\documentclass[10pt,journal,final,twocolumn,]{IEEEtran}
\usepackage{epsfig,graphicx,subfigure,psfrag,amsmath,cases,bm}
\usepackage{latexsym,amssymb,algorithm,mathtools}
\usepackage{algorithmic}
\usepackage{color}
\usepackage{url}
\usepackage{scrtime}
\usepackage{stfloats}
\usepackage{tablefootnote}
\usepackage{cite}
\usepackage{amsfonts,mathabx}
\usepackage{textcomp}
\usepackage{xcolor,capt-of}
\usepackage{multirow}
 \usepackage{tikz}
\usetikzlibrary{arrows.meta,calc,positioning,fit,shapes.geometric}
\newcommand\scalemath[2]{\scalebox{#1}{\mbox{\ensuremath{\displaystyle #2}}}}

\newtheorem{lemma}{Lemma}

\newcommand*{\rom}[1]{\expandafter\@slowromancap\romannumeral #1@}
\makeatother

\usepackage[left=0.625in,top=0.7in,bottom=0.99in,right=0.625in]{geometry}
\author{\IEEEauthorblockN {Ata Khalili, \textit{Member, IEEE}, Brikena Kaziu, \textit{Graduate Student Member, IEEE}, Vasilis K. Papanikolaou, \textit{Member, IEEE}, and Robert Schober, \textit{Fellow, IEEE}
}
\thanks{This work was supported by the Federal Ministry for Research, Technology and Space (BMFTR) in Germany in the program of ``Souverän. Digital. Vernetzt.'' joint project xG-RIC (Project-ID 16KIS2432), the Deutsche Forschungsgemeinschaft (DFG, German Research Foundation) under project GRK-2680 – Project-ID 437847244, and Horizon Europe Marie Sko\l dowska-Curie Actions (MSCA)-UNITE under project 101129618. This paper has been
presented in part at the IEEE Global Communication Conference (Globecom) 2025\cite{khalili2025pinching}. The authors are with the Institute for Digital
Communications, Friedrich-Alexander-University Erlangen–Nuremberg, 91054 Erlangen, Germany (e-mail: \{ata.khalili, brikena.kaziu, vasilis.papanikolaou, robert.schober\}(@fau.de).}}

\newtheorem{T-Prob}{Transformed Problem}

\allowdisplaybreaks

\title{\huge Clustered Movable Pinching Antennas: Realizing Beamforming Gains and Target Diversity in ISAC Systems with Look-Angle-Dependent RCS}

\begin{document}
\maketitle

%\vspace{-5mm}

\begin{abstract}
We investigate a novel integrated sensing and communication (ISAC) system enabled by pinching antennas (PAs), which are dynamically activated along a dielectric waveguide. Unlike prior designs, the PAs are organized into multiple clusters of movable antennas. The movement of the antennas within each cluster enables transmit beamforming, while the spatial separation of different clusters allows the system to illuminate the target from multiple angular perspectives. This approach provides two key benefits: (i) fine-grained beam steering within each cluster and (ii) angular diversity across clusters.
For modeling the radar echo signals, we account for the dependence of the radar cross-section (RCS) on the illumination angle. Specifically, we model the RCSs seen from different illumination angles as correlated complex Gaussian random variables, reflecting the physical continuity of the scattering behavior across nearby angular directions. We adopt the sensing outage probability as the key reliability metric, which quantifies the likelihood that the accumulated sensing signal-to-noise ratio (SNR) across time slots falls below a threshold. Accordingly, we minimize the sensing outage  probability by jointly optimizing the PA cluster selection in each time slot, the antenna positions within the activated clusters, and the cluster activation durations, subject to average data rate constraints for the communication users. The resulting optimization problem is a non-convex mixed integer non-linear program (MINLP). To obtain a tractable solution, we derive a Chernoff-bound-based surrogate for the outage probability and develop an alternating optimization algorithm within a majorization–minimization (MM) framework, where the resulting subproblems are addressed via successive convex approximation and penalty-based relaxation. Numerical results show that activating different PA clusters across time slots together with intra-cluster antenna movement for adaptive beamforming significantly reduces the sensing outage probability compared to (i) reusing the same PA cluster across time slots, (ii) fixed-position antennas, and (iii) single-antenna activation per cluster. Our results further reveal an inherent sensing–communication trade-off, where stringent communication rate requirements limit the achievable sensing diversity.
\end{abstract}
\vspace{-5mm}
\section{Introduction}

The sixth-generation (6G) wireless networks are envisioned to provide not only high-capacity and ultra-reliable communications but also real-time environmental sensing. Integrated sensing and communication (ISAC) has therefore emerged as a key enabler by allowing spectrum, hardware, and infrastructure to be shared between both functionalities. For ISAC, multiple-input multiple-output (MIMO) systems play a central role owing to their spatial processing gains that simultaneously enhance radar detection and communication reliability. However, conventional MIMO architectures rely on fixed antenna arrays with limited aperture, which fundamentally constrains adaptability and spatial resolution, particularly in dynamic and cluttered environments. To overcome these limitations, recent works have explored flexible antenna architectures such as fluid antennas and movable antennas, which can reposition their radiating elements in real time to adapt to channel variations and thereby improve system performance~\cite{wong2020fluid,zhu2023movable,khalili2025movable}.

Recently, \textit{pinching antennas} (PAs) have emerged as a promising flexible-antenna paradigm leveraging dielectric waveguides~\cite{docomo2019ntt,p1}. Unlike mechanically movable arrays, PAs enable dynamic activation of radiating elements at preconfigured locations along the waveguide without requiring physical displacement. This new architecture provides high spatial agility and scalability at low cost, while the extended aperture of the waveguide facilitates coverage over a wide area. Moreover, the low propagation loss of dielectric waveguides makes PAs particularly attractive for establishing line-of-sight (LoS) links and mitigating signal blockages in complex environments. Existing studies have demonstrated that PA-enabled systems can substantially outperform conventional fixed antenna arrays in different communication settings~\cite{p3,p4}.

While ISAC provides substantial gains in terms of spectrum and infrastructure reuse, ensuring reliable radar performance is challenging due to the angle-dependent nature of the radar cross-section (RCS). In radar theory, it is well established that the RCS varies significantly with the look angle, depending on the target's shape, material, and orientation~\cite{Radar}. Although this phenomenon has been studied in the radar community, it is often neglected for ISAC design, where sensing channels are commonly modeled as deterministic or static~\cite{ISAC6G,jsc-mimo-radar,mu-mimo-jsc}. As a result, when a target is probed from a single fixed direction, the performance may be compromised by weak echoes if the corresponding RCS happens to be small. To mitigate this limitation, distributed MIMO radar architectures have been proposed, where spatially separated antennas observe the target from multiple viewpoints, thereby achieving \textit{target diversity} and improving detection robustness~\cite{Ang1,Ang2}. Although distributed MIMO offers diversity, its implementation often requires costly and geographically separated infrastructure. A promising alternative is to exploit the flexibility of PAs, which can facilitate multi-perspective illumination of a target through dynamic activation along a dielectric waveguide. Several recent studies have investigated PA-assisted ISAC. The authors of~\cite{PASS} proposed an integrated framework for PA systems, highlighting their potential for future 6G deployments. In~\cite{PA_ISAC_sensing}, the emphasis was placed on the sensing accuracy of PA-enabled architectures, while the authors of \cite{PA_ISAC_rate} analyzed the achievable rate region of PA-assisted ISAC. More recently, the  Cramér–Rao bound (CRB) was derived to characterize the fundamental sensing limits of PA-based ISAC systems \cite{PA_ISAC_CRB}. This bound was extended to multi-waveguide architectures providing enhanced spatial diversity in \cite{PA_multi_waveguide}. However, the authors in \cite{PA_ISAC_CRB,PA_ISAC_rate,PA_ISAC_sensing,PA_multi_waveguide} consider only a single, fixed PA configuration and evaluate sensing performance within a single time slot, without accumulating sensing information across multiple time slots. Consequently, angular RCS diversity achieved through sequential activation of PAs at different positions across time slots has not been exploited in these works.

%These works collectively demonstrate the promise of PAs for ISAC; however, they do not study the sensing  reliability under angle-dependent RCSs. 

Motivated by the need to exploit angular RCS diversity for reliable sensing, the conference version of this paper \cite{khalili2025pinching}, introduced a novel PA-enabled ISAC framework where PAs at different positions are sequentially activated across time slots, allowing the target to be illuminated from multiple look angles. However, the framework in \cite{khalili2025pinching} considered only a single communication user and relied on several simplifying assumptions: (i) RCS realizations were modeled as independent across time slots, (ii) all time slots had equal duration, and (iii) only a single PA was activated in each time slot, which precluded transmit beamforming and limited the achievable performance. 

To overcome these limitations, in this paper, we extend our work in \cite{khalili2025pinching} and propose a novel PA architecture for ISAC systems, where the PAs along the waveguide are organized into several clusters. Furthermore, we model the RCSs as correlated across look angles and serve multiple users. The main contributions of this work are summarized as follows:
%Furthermore, the positions of PAs can be reconfigured within each activated cluster. 

\begin{itemize}
    \item We propose a novel ISAC architecture where multiple PA clusters are placed at fixed locations along a dielectric waveguide. Each cluster consists of multiple movable PAs that are simultaneously activated. Only one cluster is activated per time slot and the corresponding PAs are repositioned within a predefined segment along the waveguide to form a transmit beam toward the target. Different PA clusters illuminate the target with its angle-dependent RCS from different look angles, resulting in cluster-dependent RCS values, which are modeled as statistically correlated to capture the similarity of the scattering responses for nearby angles.
    \item We formulate an optimization problem that facilitates the joint design of the PA cluster and communication user selection, intra-cluster PA positioning, and cluster activation duration, with the objective to minimize the sensing outage probability, while still ensuring the quality of service (QoS) of the scheduled users. To render the problem tractable, we derive the Chernoff bound on the sensing outage probability and develop an efficient iterative solution based on the majorization minimization (MM) framework. 
   \item  Our numerical results confirm that activating different PA clusters across time slots together with intra-cluster antenna movement for adaptive beamforming significantly reduces the sensing outage probability compared to (i) reusing the same PA cluster across time slots (ii) fixed position antennas, and (iii) single-antenna activation per cluster.
 
\end{itemize}
\textit{Notation:} 
In this paper, matrices and vectors are denoted by boldface capital letters $\mathbf{A}$ and boldface lower-case letters $\mathbf{a}$, respectively. 
$\mathbf{A}^\mathrm{T}$, $\mathbf{A}^{*}$, and $\mathbf{A}^{\mathrm{H}}$ denote the transpose, complex conjugate, and Hermitian (conjugate transpose) of matrix $\mathbf{A}$, respectively. 
$\mathrm{Rank}(\mathbf{A})$, $\mathrm{Tr}(\mathbf{A})$, and $\det(\mathbf{A})$ denote the rank, trace, and determinant of $\mathbf{A}$, respectively. 
$\mathrm{diag}(\mathbf{a})$ denotes a diagonal matrix whose diagonal entries are given by the elements of vector $\mathbf{a}$, while $\mathrm{Diag}(\mathbf{A})$ denotes a vector formed by extracting the main diagonal elements of matrix $\mathbf{A}$. blkdiag $(\mathbf{A}_{1},...,\mathbf{A}_{n})$ denotes a block diagonal matrix composed of $\mathbf{A}_{1},...,\mathbf{A}_{n}$.  
$\mathbf{A}\succeq\mathbf{0}$ denotes a positive semidefinite matrix. $\mathbf{I}_N$ is the $N$-by-$N$ identity matrix. $\mathbb{R}^{N\times M}$ and $\mathbb{C}^{N\times M}$ represent the spaces of $N\times M$ real-valued and complex-valued matrices, respectively. $|\cdot|$ and $||\cdot||_2$ stand for the absolute value of a complex scalar and the $l_2$-norm of a vector, respectively. We use $\mathrm{vec}\{\cdot\}$ to represent vectorization. $\mathbf{0}_{L}$ and $\mathbf{1}_L$ represent the all-zeros and all-ones column vectors of length $L$, respectively. $\Re\{\cdot\}$ and $\Im\{\cdot\}$ denote the real and imaginary parts of a complex number,
respectively. $\mathbb{E}[\cdot]$ refers to statistical expectation. 
\begin{figure}
    \centering
    \includegraphics[ width=0.95\linewidth]{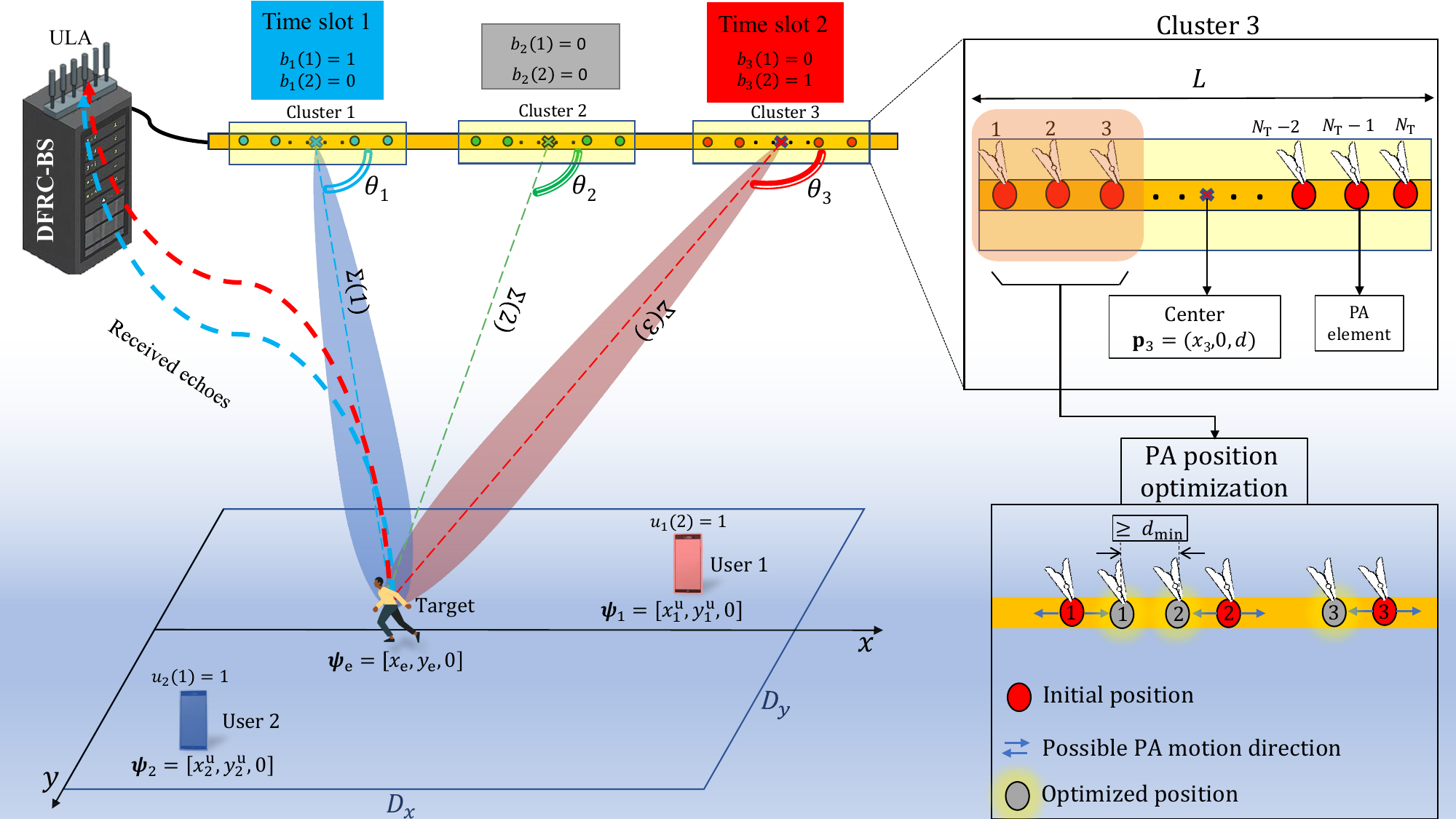}
    \caption{Illustration of the considered downlink PA-assisted ISAC system with a DFRC-BS, dynamically activated PA clusters, multiple communication users, a radar receiver, and a target. $\Sigma(m)$ denotes the RCS coefficient for cluster $m$, which depends on the corresponding look angle $\theta_{m}$.}
    \label{fig:PA_system_model}
\end{figure}
\vspace{-4mm}
\section{System Model}
We consider a downlink PA-enabled ISAC system comprising a dual-function radar-communication (DFRC) base station (BS), equipped with a PA array mounted on a dielectric waveguide, as illustrated in Fig.~\ref{fig:PA_system_model}. The dielectric waveguide of length $D_x$ is deployed along the $x$-axis at a fixed height $d$ above the ground. The PA elements along the waveguide are organized into \( M \) spatially separated clusters, each comprising a group of \( N_\mathrm{T} \) movable PAs. The center of the $m$-th PA cluster, $m \in \mathcal{M} =  \{1, \ldots,M\}$, is preconfigured at \( \mathbf{p}_m = [x_m, 0, d] \in \mathbb{R}^3 \), and \( x_m \in [0, D_x] \). 

During each time slot $t$, $t \in \mathcal{T} = \{1, \ldots, T\}$, only one PA cluster is selected to serve one user $k$, $k \in$  \( \mathcal{K} = \{1, \dots, K\} \), for downlink communication, while simultaneously probing a single static target. Time slot $t$ is assigned a duration \( \tau(t) \in [0, T_{\max}] \), and the total time allocated across all slots is constrained as follows:
\begin{equation}
    \sum_{t=1}^{T} \tau(t) \leq T_{\max}.
\end{equation}
User positions $\boldsymbol{\psi}_k = [x_k^{\mathrm{u}}, y_k^{\mathrm{u}}, 0] \in \mathbb{R}^3$, $k \in \mathcal{K}$, and target position  \( \boldsymbol{\psi}_\mathrm{e} = [x_\mathrm{e}, y_\mathrm{e}, 0]\in \mathbb{R}^3 \) are assumed to be fixed during one transmission frame and are known at the DFRC-BS. Selecting  different PA clusters in different time slots provides different angular views $\theta_m$, $m \in \mathcal{M}$, of the target over time, see Fig.~\ref{fig:PA_system_model}. These different angular perspectives introduce \textit{target diversity}, which is critical for overcoming the randomness and directionality of the RCS. To be able to effectively suppress self-interference and enhance sensing performance, we assume that the radar echoes are collected at the DFRC-BS using a separate, fixed uniform linear array (ULA) equipped with \( N_{\mathrm{R}} \) antennas.
\vspace{-2mm}
\subsection{Pinching Antenna Configuration}
In the proposed ISAC system, the transmit array is composed of multiple clusters of movable PAs distributed along a dielectric waveguide. Each cluster contains a small number of PAs whose physical positions can be dynamically adjusted within the cluster. This architecture is intentionally designed to strike a balance between performance and practical deployability. While moving all antennas freely across the entire waveguide would maximize the spatial degrees of freedom (DoFs), such a configuration would incur prohibitive mechanical complexity, high latency, and susceptibility to positioning errors. Instead, we restrict antenna movement to within pre-designed cluster regions. This allows each cluster to independently perform flexible beam steering with low reconfiguration latency, enabling rapid adaptation to sensing and communication requirements without the need for coordinated movement across the entire waveguide.

The \( N_\mathrm{T} \) PAs of each cluster can slide continuously within a short linear aperture of length \( L \), centered at position \( x_m \).
This reconfigurable structure enables physical adjustment of antenna locations, offering fine spatial control over the transmitted wavefront and enabling beam steering without digital precoding. In each time slot \( t \in \mathcal{T} \), a single cluster is activated to transmit the joint sensing and communication signal. 
Let \( b_m(t) \in \{0,1\} \) denote the corresponding binary activation variable, satisfying
\begin{equation}
    \sum_{m=1}^{M} b_m(t) = 1, \quad \forall t \in \mathcal{T}.
\end{equation}
The set of cluster centers \(x_m\) is predesigned to ensure sufficiently distinct look angles toward the target, ensuring that each cluster provides a unique spatial perspective. Within the active cluster, the positions of the movable antennas are optimized to form directional beams toward the communication user and the sensing target. 
Let \( x_{m,n} \in [x_m - L/2,\, x_m + L/2] \) denote the position of the \( n \)-th antenna of cluster \( m \), where \( n \in \{1, \dots, N_\mathrm{T}\} \). 
To avoid mutual coupling and physical overlap, the positions of the movable PAs must satisfy a minimum spacing constraint, ensuring that the distance between any pair of PAs remains above a predefined threshold as follows:
\begin{equation}
    x_{m,n+1} - x_{m,n} \geq d_{\min}, \quad \forall n,~\forall m.
\end{equation}
By jointly optimizing the discrete cluster selection variables \( b_m(t) \) and the continuous intra-cluster antenna positions \( x_{m,n} \), the proposed PA configuration can provide coarse angular diversity across clusters and fine-grained beamforming within each cluster, benefiting both communication and sensing performance.

\subsection{Transmit Signal Model}
In each time slot \( t \), to maximize spectral efficiency, the DFRC-BS transmits a single complex baseband waveform that simultaneously carries communication data and performs radar probing. The transmitted signal is given by
\begin{equation}
    x(t) = \sqrt{p_\mathrm{T}} \, c(t),
\end{equation}
where \( c(t) \in \mathbb{C} \) is the information-bearing symbol intended for the user selected in time slot $t$, satisfying \( \mathbb{E}[|c(t)|^2] = 1 \), $\forall t$, and $p_\mathrm{T}\in \mathbb{R}^+$ is the transmit power of the BS. Signal $c(t)$ is injected into the waveguide at the feed point \( \mathbf{p}_0 = [x_0, 0, d] \).

\subsection{Communication Channel and Metric}
In the proposed PA-enabled ISAC system, the communication channel is determined by both the free-space propagation from the active PAs to the users and the guided propagation inside the dielectric waveguide. 
As illustrated in Fig.~\ref{fig:PA_system_model}, each PA  radiates the signal launched from the feed point $\mathbf{p}_0$, where the signal first travels through the waveguide to the PA and then propagates through free space to the communication users. 
Hence, the composite channel consists of two physically distinct parts:

\begin{itemize}
    \item \textit{Waveguide propagation:} The signal experiences attenuation and phase shift while traveling a guided distance of $\ell_{m,n} = \|\mathbf{p}_0 - \mathbf{x}_{m,n}\|$ from the feed point to the $n$-th PA element of cluster $m$, where $\mathbf{x}_{m,n} = [x_{m,n},\,0,\,d]$.
    The amplitude loss is modeled as $e^{-\alpha \ell_{m,n}}$, where $\alpha$ is the waveguide attenuation coefficient, and the guided phase term $e^{-j\frac{2\pi}{\lambda_g}\ell_{m,n}}$ accounts for propagation with guided wavelength $\lambda_g = \lambda/n_\text{eff}$, where $\lambda$ and $n_\text{eff}$ are the free-space wavelength and the effective refractive index of the dielectric material, respectively.
    
    \item \textit{Free-space propagation:} After radiation from the $n$-th PA of cluster $m$, the signal travels distance $d_{k,m,n} = \|\boldsymbol{\psi}_k - \mathbf{x}_{m,n}\|$ through the air to user $k$, experiencing joint free-space attenuation and phase rotation given by $\frac{1}{d_{k,m,n}} e^{-j\frac{2\pi}{\lambda}d_{k,m,n}}$.
\end{itemize}

By combining these two propagation effects, the overall complex baseband channel coefficient from the feed point to user \( k \), $k \in \mathcal{K}$, through the $n$-th PA of cluster $m$, is given by
\begin{equation}\label{eq:commun_channel}
h_{k,m,n} =
\frac{\eta}{d_{k,m,n}\sqrt{N_\mathrm{T}}}
e^{-\alpha \ell_{m,n}}
e^{-j\left(\frac{2\pi}{\lambda_g}\ell_{m,n} + \frac{2\pi}{\lambda}d_{k,m,n}\right)},
\end{equation}
where \( \eta = \frac{c}{4\pi f_c} \) is the free-space pathloss constant, $c$ is the speed of light, and $f_c$ is the carrier frequency. The normalization factor $\frac{1}{\sqrt{N_{\mathrm{T}}}}$ ensures that the total transmit power remains constant independent of the number of PAs per cluster, such that any performance gain results solely from coherent combining rather than power scaling. Eq.~\eqref{eq:commun_channel} reveals that changing the PA position $\mathbf{x}_{m,n}$ simultaneously affects the guided phase, and the free-space phase, and thus the overall channel phase. 
This dependency of the phase on the PA position is fundamental to the beamforming capability of movable PAs. 

Next, let us define the binary scheduling variable \( u_k(t) \in \{0,1\} \), where \( u_k(t) = 1 \) indicates that user \( k \) is served in time slot \( t \). To ensure that at most one  user is selected in each time slot, the following constraint is imposed:
\begin{equation}
    \sum_{k=1}^{K} u_k(t) \leq  1. \quad 
\end{equation}
Then, the received signal at user $k$ is given by 
\begin{equation}
y_k(t) = \sum_{m=1}^{M} b_m(t) \sum_{n=1}^{N_\mathrm{T}} h_{k,m,n} \sqrt{p_\mathrm{T}}\, c(t) + z_k(t),
\end{equation}
where \( z_k(t) \sim \mathcal{CN}(0, \sigma_{k}^2) \) is additive white Gaussian noise (AWGN) with variance $\sigma_{k}^2$. Consequently, the instantaneous received SNR $\gamma_k(t)$ at scheduled user \( k \), i.e.,  for $u_k(t) = 1$, is given as follows:
\begin{equation}
\gamma_k(t) = \frac{p_\mathrm{T}}{\sigma_{k}^2} \left| \sum_{m=1}^{M} b_m(t) \sum_{n=1}^{N_\mathrm{T}} h_{k,m,n} \right|^2.
\end{equation}

\subsection{Sensing Channel and Metric}
The sensing channel is affected by the same physical propagation phenomena as the communication channel, but also includes the reflection from the target. 
Specifically, the transmitted signal propagates along the dielectric waveguide from the feed point to the active PA element, is radiated toward the target, and the reflected echo is collected by the fixed receive ULA at the DFRC-BS. 
Accounting for both guided and free-space propagation, the baseband sensing channel coefficient between the $n$-th PA in cluster \( m \) and the target is given by
\begin{align}\label{eq:sensing_chann_coeff}
h_{\mathrm{e},m,n} =&
\frac{\eta}{\|\boldsymbol{\psi}_\mathrm{e} - \mathbf{x}_{m,n}\|\sqrt{N_\mathrm{T}}} 
e^{-\alpha \|\mathbf{p}_0 - \mathbf{x}_{m,n}\|}\nonumber\\& 
e^{-j \left( \frac{2\pi}{\lambda_g} \|\mathbf{p}_0 - \mathbf{x}_{m,n}\| + \frac{2\pi}{\lambda} \|\boldsymbol{\psi}_\mathrm{e} - \mathbf{x}_{m,n}\| \right)}.
\end{align}

In the following, we introduce the correlated RCS model and derive the corresponding accumulated radar SNR and outage probability for the system model shown in Fig. \ref{fig:PA_system_model} and the channel model given in \eqref{eq:sensing_chann_coeff}.  

%This channel represents the deterministic propagation component from the feed point to the target. The overall radar echo also includes the target’s radar cross-section (RCS), which is modeled as a random, look-angle-dependent process and will be incorporated in the radar signal model and performance metric.
\subsubsection*{1) Correlated RCS Modeling}
In our system, the spatially distributed PA clusters illuminate the target from different angular directions. Specifically, each PA cluster $m \in \mathcal{M}$ probes the target from a unique angle of incidence $\theta_m$, as illustrated in Fig.~\ref{fig:PA_system_model}. Since the RCS of the target is highly dependent on the angle from which it is illuminated, each cluster experiences a different RCS. In contrast, the antennas within a single cluster are confined to a small spatial region of length $L$ along the waveguide. Since $L$ is small compared to the distance between the PA cluster and the target, the variation in the illumination angle across the PAs within the same cluster is negligible. As a result, all PAs within cluster $m$ are assumed to illuminate the target from approximately the same look angle $\theta_{m}$ and therefore experience the same effective RCS, denoted by $\Sigma(m)$. To capture the spatial variations of the RCS across different clusters, we model the RCS for the $M$ clusters as a random vector \cite{Radar}:
\begin{equation}\label{eq:RCS_vec}
\boldsymbol{\Sigma}_\mathrm{e} = [\Sigma(1), \Sigma(2), \dots, \Sigma(M)]^\mathrm{T} \sim \mathcal{CN}(\mathbf{0}, \mathbf{R}_\Sigma),
\end{equation}
where \( \mathbf{R}_\Sigma \in \mathbb{C}^{M \times M} \) is the covariance matrix that reflects the angular correlation between different RCSs. Following established models in the literature \cite{Exp,richards2014fundamentals,Correlation}, we use an exponential function to model the correlation between $\Sigma(i)$ and $\Sigma(j)$, $i, j \in \mathcal{M}$:
\begin{equation}
[\mathbf{R}_\Sigma]_{i,j} = \zeta_{\mathrm{av}} e^{-\kappa |\theta_i - \theta_j|}, \quad i,j \in \mathcal{M},
\end{equation}
where $\zeta_{\mathrm{av}}=\mathbb{E}\{\|\boldsymbol{\Sigma}_\mathrm{e}\|^{2}\}$ is the average RCS power gain and \( \kappa \geq 0 \) denotes the angular correlation decay rate that controls the degree of statistical similarity between the RCSs corresponding to different look angles. The resulting round-trip radar channel between the BS and the target can be expressed as follows:
\begin{equation}
\mathbf{g}(t) = \frac{ \mathbf{a}_\mathrm{r}(\theta_\mathrm{e})}{d_{\mathrm{e},\mathrm{r}}}  \, \sum_{m=1}^{M} \Sigma(m)b_m(t) \sum_{n=1}^{N_\mathrm{T}} h_{\mathrm{e},m,n},
\end{equation}
where \( d_{\mathrm{e},\mathrm{r}} \) is the distance between the target and the receive array, \( \mathbf{a}_\mathrm{r}(\theta_\mathrm{e}) \in \mathbb{C}^{N_\mathrm{R}} \) is the receive steering vector aligned with the known direction of the target $\theta_{\mathrm{e}}$ \cite{khalili2024efficient}.

The received echo at the DFRC-BS can be expressed as 
\begin{equation}
\mathbf{r}_\mathrm{e}(t) = \mathbf{g}(t)x(t) + \mathbf{z}(t),
\end{equation}
where \( \mathbf{z}(t) \sim \mathcal{CN}(\mathbf{0}, \sigma^2 \mathbf{I}_{N_\mathrm{R}}) \) is the AWGN vector at the receive ULA with variance of $\sigma^2$. The DFRC-BS applies a matched receive beamformer which is given as follows:
\begin{equation}
\mathbf{v} = \frac{\mathbf{a}_\mathrm{r}(\theta_\mathrm{e})}{\|\mathbf{a}_\mathrm{r}(\theta_\mathrm{e})\|},
\end{equation}
and the resulting received radar signal is given by
\begin{equation}
\tilde{r}_\mathrm{e}(t) = \mathbf{v}^\mathrm{H} \mathbf{r}_\mathrm{e}(t) = \mathbf{v}^\mathrm{H} \mathbf{g}(t) x(t) + \mathbf{v}^\mathrm{H} \mathbf{z}(t).
\end{equation}
Then, the normalized instantaneous sensing SNR in time slot \( t \), i.e., $\Gamma(t)$, $\forall t$, can be expressed as:
\begin{equation}\label{eq:gamma_t}
\Gamma(t) = \frac{\tau(t)}{T_{\max}} \psi\, \left|\sum_{m=1}^{M} b_m(t)\Sigma(m) \sum_{n=1}^{N_\mathrm{T}} h_{\mathrm{e},m,n}\right|^2,
\end{equation}
where the deterministic gain $\psi$ is given by
\begin{equation}
\psi = \frac{p_\mathrm{T} \beta_0^2}{\sigma^2 d_{\mathrm{e},\mathrm{r}}^2}  \|\mathbf{a}_\mathrm{r}(\theta_\mathrm{e})\|^2.
\end{equation}

Note that in narrowband radar systems operating with constant transmit power $p_{\mathrm{T}}$, the received signal energy and hence the effective SNR scale linearly with observation time $\tau(t)$, $\forall t$. Therefore, the inclusion of \( \tau(t) \) in the SNR expression in \eqref{eq:gamma_t} captures this physical reality where the radar echo strength increases proportionally with the time duration during which the target is illuminated \cite{khalili2025movable}. By normalizing with respect to the total time budget \( T_{\max} \) (see \eqref{eq:gamma_t}), we preserve unit consistency and ensure that \( \Gamma(t) \) represents the per-slot contribution to the overall accumulated sensing SNR.

\subsubsection*{2) Accumulated Sensing SNR and Outage Probability}

To assess the sensing performance over the entire transmission period \(T_{\max}\), we consider the accumulated sensing SNR obtained by combining the sensing observations across all time slots. Specifically, the accumulated sensing SNR is given by
\begin{equation}\label{eq:gamma_acc}
\Gamma_{\mathrm{acc}}
= \sum_{t=1}^{T} \Gamma(t)
= \psi\sum_{t=1}^{T} \frac{\tau(t)}{T_{\max}} 
\left|
\sum_{m=1}^{M}
b_m(t)\,\Sigma(m)
\sum_{n=1}^{N_\mathrm{T}} h_{\mathrm{e},m,n}
\right|^2 .
\end{equation}

Since the cluster selection variables satisfy \(b_m(t)\in\{0,1\}\) and \(\sum_{m=1}^{M} b_m(t)= 1\), only one PA cluster is active in each time slot. As a result, the instantaneous sensing SNR can be simplified to
\begin{equation}\label{eq:gamma_simplified}
\Gamma(t)
=
\psi\sum_{m=1}^{M}
\frac{\tau(t)}{T_{\max}}
\,
b_m(t)
\left|
\sum_{n=1}^{N_\mathrm{T}} h_{\mathrm{e},m,n}
\right|^2
|\Sigma(m)|^2 .
\end{equation}

Then, the total sensing SNR in \eqref{eq:gamma_acc} can be expressed as
\begin{equation}\label{eq:gamma_acc_simplified}
\Gamma_{\mathrm{acc}}
=
\sum_{m=1}^{M}
q_m\,|\Sigma(m)|^2 ,
\end{equation}
where the deterministic weight associated with cluster \(m\) is defined as
\begin{equation}
q_m
\triangleq
\psi\sum_{t=1}^{T}
\frac{\tau(t)}{T_{\max}}
\,
b_m(t)
\left|
\sum_{n=1}^{N_\mathrm{T}} h_{\mathrm{e},m,n}
\right|^2 .
\end{equation}

 Since $\Sigma(m)$ is modeled as a correlated complex Gaussian random variable, see \eqref{eq:RCS_vec}, $\left|\Sigma(m)\right|^2$ follows a correlated exponential distribution. As a result, $\Gamma_{\text {acc }}$ is a weighted sum of correlated non-identically exponentially distributed random variables, see \eqref{eq:gamma_acc_simplified}. To model the reliability of the sensing process, we adopt the \textit{sensing outage  probability} as performance metric, defined as:
\begin{equation}\label{eq:out_prob_gen_formula}
P_{\mathrm{out}} = \Pr\left( \Gamma_{\mathrm{acc}} < \Gamma^{\mathrm{th}} \right),
\end{equation} where \( \Gamma^{\mathrm{th}} \) denotes the minimum SNR threshold required for successful target detection. Given the statistical uncertainty induced by the randomness of the RCS, the probabilistic formulation in \eqref{eq:out_prob_gen_formula} provides a rigorous metric for assessing sensing performance while explicitly accounting for the RCS correlations across different PA clusters.
\section{Problem Formulation}

We aim to jointly optimize antenna cluster selection, user scheduling, intra-cluster antenna positioning, and time allocation for minimization of the sensing outage probability, while the long-term communication QoS must be guaranteed for the scheduled users. The sensing outage probability captures the likelihood that the sensing SNR accumulated across all time slots falls below detection threshold \( \Gamma^{\mathrm{th}} \), resulting in unreliable target detection. The randomness of the RCS introduces stochastic uncertainty regarding the received echoes. By activating different spatially separated PA clusters along the waveguide in different time slots and adjusting the antenna positions within each activated cluster, angular diversity and a beamforming gain are realized, respectively, which improves sensing reliability. The resulting optimization problem is formulated as follows
\begin{align}
\mathcal{P}_{0}: \quad 
&\underset{\substack{\{b_m(t),~\tau(t),~u_k(t),~x_{m,n}\}}}{\text{minimize}} \quad 
\mathcal{F}\triangleq\Pr\left( \Gamma_{\mathrm{acc}} < \Gamma^{\mathrm{th}} \right) \nonumber \\
\text{s.t.} \quad
&\text{C1: }\sum_{t=1}^{T} \frac{\tau(t)}{T_{\max}} u_k(t)\, \log_2\left(1 + \gamma_k(t)\right) \geq R_{\min}, \forall k \in \mathcal{K}, \nonumber\\
&\text{C2:} \sum_{m=1}^{M} b_m(t) = 1, \quad \forall t \in \mathcal{T}, \nonumber\\
&\text{C3: }\sum_{t=1}^{T} \tau(t) \leq T_{\max}, \quad \tau(t) \geq T_{\min}, \quad \forall t \in \mathcal{T},  \nonumber\\
&\text{C4: } b_m(t) \in \{0,1\}, \quad \forall m \in \mathcal{M},\ t \in \mathcal{T}, \nonumber\\
&\text{C5: }u_k(t) \in \{0,1\}, \quad \forall k \in \mathcal{K},\ t \in \mathcal{T}, \nonumber \\
&\text{C6: }\sum_{k=1}^{K} u_k(t) \leq 1, \quad \forall t \in \mathcal{T}, \nonumber\\
&\text{C7: }x_{m,n+1} - x_{m,n} \geq d_{\min}, \quad \forall n, \ \forall m \in \mathcal{M}, 
\end{align}
where C1 ensures that the average communication rate meets the required minimum data rate \( R_{\min} \). C2 enforces that only one cluster is activated in each time slot $t$. Constraint~C3 ensures that the total time allocated across all time slots does not exceed the overall budget \( T_{\max} \). At the same time, each individual time slot duration \( \tau(t) \) must be no smaller than a minimum threshold \( T_{\min} \), which accounts for hardware and processing limitations (e.g., minimum actuation or integration time). C4 and C5 guarantee binary cluster selection and user scheduling decisions, respectively. C6 ensures that at most one user is served in each time slot. Finally,~C7 enforces a minimum separation \( d_{\min} \) between any two PAs within each cluster to mitigate mutual coupling and ensure stable beamforming.

Problem \( \mathcal{P}_0 \) is non-convex due to the probabilistic objective function, the binary cluster selection variables, the binary user scheduling variables, and the coupling between antenna positioning and time-slot allocation. In the next section, we derive a tractable solution for \( \mathcal{P}_0 \) based on a Chernoff-bound approximation and the MM approach.

\section{Proposed Solution}
Direct evaluation of the sensing outage  probability is challenging because the accumulated sensing SNR $\Gamma_\mathrm{acc}$ is a weighted sum of statistically dependent RCS power terms, $|\Sigma(m)|^{2}$, $\forall m$, see \eqref{eq:gamma_acc_simplified}. %This dependence arises from the correlated complex Gaussian modeling of the RCS vector in \eqref{eq:RCS_vec}.
To enable a tractable analysis, let us first express the accumulated sensing SNR $\Gamma_{\mathrm{acc}}$ as a  quadratic form. Specifically, by defining
\( \mathbf{Q} = \mathrm{diag}(q_1,\ldots,q_M) \),
the accumulated sensing SNR can be compactly written as
\begin{equation}
\Gamma_{\mathrm{acc}}
=
\boldsymbol{\Sigma}_\mathrm{e}^{\mathrm{H}}
\mathbf{Q}
\boldsymbol{\Sigma}_\mathrm{e}.
\end{equation}

This representation enables the application of a Chernoff-bound-based approximation, which yields an analytically tractable upper bound on the sensing outage  probability while explicitly accounting for the angle-dependent RCS statistics. 
Using the Chernoff bound for complex Gaussian quadratic forms \cite{chernoff,khalili2025pinching}, the outage probability can be upper-bounded as follows:
\begin{equation} \label{chernoff}
P_{\mathrm{out}} \leq \min_{s > 0} \left\{ e^{s \Gamma^{\mathrm{th}}} \frac{1}{\det\left( \mathbf{I}_{M} + s \mathbf{R}_\Sigma \mathbf{Q} \right)} \right\}
\end{equation}

Taking the logarithm of the argument in \eqref{chernoff}  yields the following surrogate objective function:
\begin{equation}
\min_{s > 0} \left\{ s \Gamma^{\mathrm{th}} -\log \det\left( \mathbf{I}_{M} + s \mathbf{R}_\Sigma \mathbf{Q} \right) \right\}.
\end{equation}

In the following, for a fixed Chernoff parameter 
$s$, the resulting surrogate problem is solved using an alternating optimization (AO) framework. In this framework, the surrogate problem is partitioned into two tractable subproblems that are optimized iteratively until convergence. Specifically, the first subproblem optimizes the PA cluster selection and time slot duration, while the second subproblem optimizes the intra-cluster antenna positions and performs communication user scheduling. The continuous subproblems are addressed within an MM framework implemented via successive convex approximation (SCA), while the discrete variables are handled using big-M relaxation. Since the Chernoff-bound-based surrogate depends on the scalar parameter $s>0$, we perform a one-dimensional search over a finite grid of candidate values. For each candidate $s$, the AO procedure is executed until convergence, and the solution yielding the smallest surrogate value is selected. The overall solution procedure, including the outer search over $s$ and the inner alternating updates, is summarized in Fig.~2.

\begin{figure}[t!] % Placement 
\centering
\includegraphics[width=.88\linewidth]{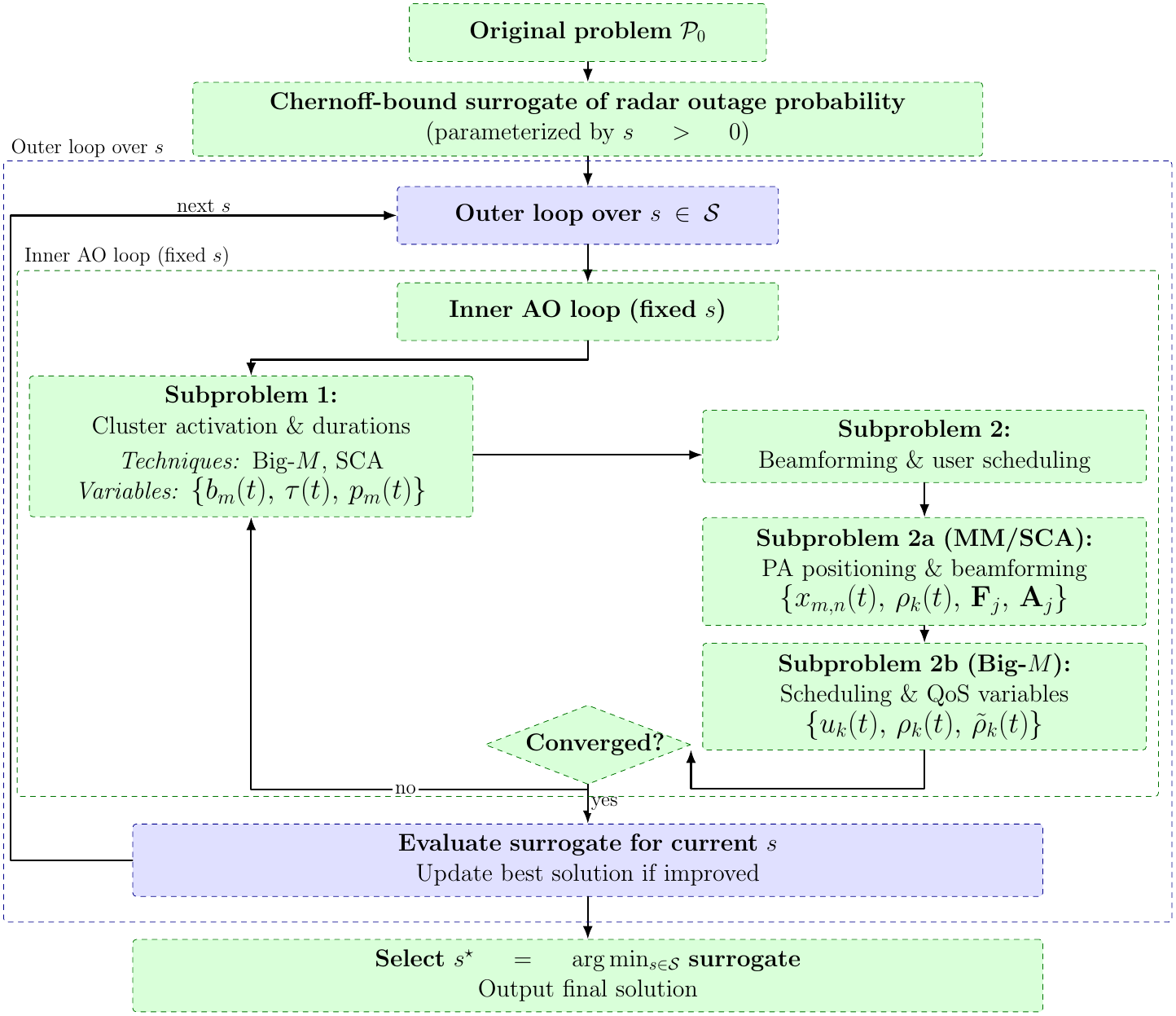}
        %\vspace{-55mm}
	 	\caption{ \small Block diagram of the proposed solution to problem $\mathcal{P}_{0}$ based on the AO-based Algorithm 1.}\label{blockdiagram}
\end{figure}

\subsection{Subproblem 1: Cluster Selection and Activation Duration Optimization}

In the first subproblem, for given PA positions and scheduled users, we jointly optimize the binary cluster selection variables $b_m(t)$ and activation durations $\tau(t)$ for all time slots. The problem can be formulated as follows
\begin{align}
\mathcal{P}_1: \quad & \underset{{b_m(t), \tau(t)}}{\text{minimize}} \quad  s\Gamma^{\mathrm{th}} - \log \det\left( \mathbf{I}_{M} + s\mathbf{R}_\Sigma \mathbf{Q} \right)\\
\text{s.t.} \quad
&\text{C1},\text{C2},\text{C3},\text{C4}. \nonumber
\end{align}
The above problem is still non-convex. The non-convexity arises from two sources: (i) the log-determinant term, which depends on the binary variables $b_{m}(t)$ through $\mathbf{Q}$, and (ii) the bilinear terms $\tau(t)b_m(t)$. To address (i), we apply a first-order Taylor approximation in SCA iteration $l$. We define
\[
\mathbf{A}^{(l)} \triangleq \mathbf{I}_M + s\, \mathbf{R}_\Sigma \mathbf{Q}^{(l)},
\]
then
\begin{align}
\log \det\left( \mathbf{I}_M + s\, \mathbf{R}_\Sigma \mathbf{Q} \right)
&\leq \log \det\left( \mathbf{A}^{(l)} \right) +\nonumber\\& 
\mathrm{Tr} \left( \left(\mathbf{A}^{(l)}\right)^{-1} s\, \mathbf{R}_\Sigma (\mathbf{Q} - \mathbf{Q}^{(l)}) \right).
\end{align}

To handle the bilinear term \( \tau(t) b_m(t) \), we introduce auxiliary variables:
\begin{equation}
p_m(t) \triangleq \tau(t) b_m(t), \quad \forall m, t,
\end{equation}
and impose the following big-M constraints:
\begin{align}
& \text{C8}: 0 \leq p_m(t) \leq T_{\max} b_m(t), \quad \forall m, t, \\
& \text{C9}: \tau(t) - (1 - b_m(t)) T_{\max} \leq p_m(t) \leq \tau(t), \quad \forall m, t.
\end{align}

Next, we relax the binary constraint \( b_m(t) \in \{0,1\} \) and rewrite C4 as follows: 
\begin{align}
& \text{C4a}: 0\leq b_{m}(t)\leq 1,\\
& \text{C4b}: \sum_{m=1}^{M} \sum_{t=1}^{T} \left( b_m(t) - \big(b_m(t)\big)^{2} \right)\leq 0. 
\end{align}
Constraint C4b is a difference of convex (DC) functions, and we use the first-order Taylor expansion to convert this non-convex constraint to the following convex constraint 
\begin{equation}
    \scalemath{0.9}{\widehat{\text{C4b}}: \sum_{m=1}^{M} \sum_{t=1}^{T} \left( b_m(t) - b_m^{(l)}(t) \right) \left( 2 b_m^{(l)}(t) - b_m(t) \right) \leq 0.}
\end{equation}
Now, we introduce a penalty factor $\rho$ to move constraint C4b
to the objective function, where $\rho$ represents the relative importance
of recovering binary values for $b_{m}(t)$. For a sufficiently large
value of $\rho$, optimization problem $\mathcal{P}_{1}$ is equivalent to the
following convexified Subproblem 1 in iteration \( l \):
\begin{align}\label{P1}
\mathcal{P}_1^{(l)}: \quad & \underset{\substack{b_m(t), \tau(t), p_m(t)}}{\text{minimize}} \quad 
s \Gamma^{\mathrm{th}} - \log \det\left( \mathbf{A}^{(l)} \right) -\nonumber\\& 
\mathrm{Tr} \left( \left(\mathbf{A}^{(l)}\right)^{-1} s\, \mathbf{R}_\Sigma (\mathbf{Q} - \mathbf{Q}^{(l)}) \right)+\nonumber\\&\rho \sum_{m=1}^{M} \sum_{t=1}^{T} \left( b_m(t) - b_m^{(l)}(t) \right) \left( 2 b_m^{(l)}(t) - b_m(t) \right) \nonumber \\
\text{s.t.} \quad
&\text{C1: } \sum_{t=1}^{T} \frac{p_m(t)}{T_{\max}} u_k(t) \log_2(1 + \gamma_{k}(t)) \geq R_{\min},\forall k, \nonumber \\
& \text{C2: } \sum_{m=1}^{M} b_m(t) = 1, \quad \forall t, \nonumber \\
& \text{C3: } \sum_{t=1}^{T} \tau(t) \leq T_{\max}, \quad \tau(t) \geq T_{\min}, \quad \forall t, \nonumber \\
& \text{C4a: } 0 \leq b_m(t) \leq 1, \quad \forall m, t, \nonumber \\
& \text{C8-C9}. 
\end{align}
This subproblem is convex and can be optimally solved via known convex solvers such as CVX\cite{CVX}.
\subsection{Subproblem 2: Beamforming via Movable PAs
and User Scheduling}

Given the selected PA clusters and activation times from Subproblem~1, we next optimize the intra-cluster antenna positions ${x}_{m,n}$ and the binary user scheduling variables $u_k(t)$. This subproblem is formulated as follows:
\begin{align}
\mathcal{P}_2: \quad 
& \underset{{u_k(t), \,{x}_{m,n}}}{\text{minimize}}
\ \ \ s\,\Gamma^{\mathrm{th}} - \log \det\!\left( \mathbf{I}_M + s\,\mathbf{R}_{\Sigma} \,\mathbf{Q}\right) \nonumber\\
\text{s.t.} \quad
&\text{C1}, \text{C5}, \text{C6}, \text{C7}.
\end{align}

The free-space distance between the $n$-th PA and a generic node $o\in \{k,e\}$, where $k$ and $e$ index the communication users and the target, respectively, is given by
\begin{equation}
d_{o,m,n}= \big| \boldsymbol{\psi}_o - \mathbf{x}_{m,n} \big|
= \sqrt{\big(x_{m,n} - x_o\big)^2 + y_o^2 + d^2},
\end{equation}
and the waveguide feed-to-antenna distance is
$\ell_{m,n}$. 
For analytical convenience, we separate the amplitude and phase components of the channel. Specifically, the channel between the $n$-th antenna of cluster $m$ and user $k$ is decomposed as
$f_{m,n,k} \triangleq \frac{1}{d_{k,m,n}}$
and
$a_{m,n,k} \triangleq e^{-j\theta_{m,n,k}}$,
where the phase is given by
$\theta_{m,n,k} = \frac{2\pi}{\lambda} d_{k,m,n} + \frac{2\pi}{\lambda_g} \ell_{m,n}$. For each cluster, define the vectors
$\mathbf f_{m,k} = [f_{m,1,k}, \ldots, f_{m,N_\mathrm{T},k}]^{\mathrm{T}} \in \mathbb{R}^{N_\mathrm{T}}$
and
$\mathbf a_{m,k} = [a_{m,1,k}, \ldots, a_{m,N_\mathrm{T},k}]^{\mathrm{T}} \in \mathbb{C}^{N_\mathrm{T}}$,
and matrix
$\mathbf C_m
=
\mathrm{diag}(\eta e^{-\alpha \ell_{m,1}}, \ldots, \eta e^{-\alpha \ell_{m,N_\mathrm{T}}})$. Stacking across clusters yields
$\mathbf f_k = \mathrm{vec}\{\mathbf f_{1,k}, \ldots, \mathbf f_{M,k}\}$,
$\mathbf a_k = \mathrm{vec}\{\mathbf a_{1,k}, \ldots, \mathbf a_{M,k}\}$,
and
$\mathbf C
=
\mathrm{blkdiag}(b_{1}(t)\mathbf C_1, \ldots, b_{M}(t)\mathbf C_M)$.

We further define
$\mathbf F_k = \mathbf f_k \mathbf f_k^{\mathrm{H}}$
and
$\mathbf A_k = \mathbf a_k \mathbf a_k^{\mathrm{H}}$. Besides, we also introduce the auxiliary variable $\rho_k(t)$ to bound the SNR for the communication user as follows
\begin{equation}
    \frac{p_{\mathrm{T}}}{\sigma_k^2}
\left|
\sum_{m=1}^{M}
b_m(t)
\sum_{n=1}^{N_{\mathrm{T}}}
h_{k,m,n}
\right|^2
\ge \rho_k(t)\geq 0.
\end{equation}
Since $b_m(t)\in\{0,1\}$ and $\sum_{m=1}^M b_m(t)=1$, exactly one cluster is active in each time slot. Hence, the term $\sum_{m=1}^{M} b_m(t) h_{k,m,n}$ selects the channel coefficient corresponding to the active cluster only. Using the matrix representation of the effective channel, we have
\[
\left|
\sum_{n=1}^{N_\mathrm{T}}
\sum_{m=1}^{M}
b_m(t) h_{k,m,n}
\right|^2
=
\mathrm{Tr}
\big(
\mathbf C^{\mathrm{H}}
\mathbf A_k
\mathbf C
\mathbf F_k
\big).
\]
To obtain a tractable reformulation, we use the Frobenius identity
$\|\mathbf X + \mathbf Z\|_{\mathrm{F}}^2
=
\|\mathbf X\|_{\mathrm{F}}^2
+
\|\mathbf Z\|_{\mathrm{F}}^2
+
2\Re\{\mathrm{Tr}(\mathbf X^{\mathrm{H}} \mathbf Z)\}$. Rearranging this identity gives 
$\Re\{\mathrm{Tr}(\mathbf X^{\mathrm{H}} \mathbf Z)\}=\frac{1}{2}\big(\|\mathbf X + \mathbf Z\|_{\mathrm{F}}^2-\|\mathbf X\|_{\mathrm{F}}^2-\|\mathbf Z\|_{\mathrm{F}}^2\big)$. Applying this result with $\mathbf{X}=\mathbf{F}_{k}$ and $\mathbf{Z}=\mathbf C^{\mathrm{H}}
\mathbf A_k \mathbf C$, the SNR for the communication users can be written as 
\begin{equation}
  \text{C1a:}~\|\mathbf F_k+\mathbf C^{\mathrm{H}}\mathbf A_k\mathbf{C}\|_{\mathrm{F}}^2
-
\|\mathbf F_k\|_{\mathrm{F}}^2
-
\|\mathbf C^{\mathrm{H}}\mathbf A_k\mathbf C\|_{\mathrm{F}}^2
\ge
2\frac{\sigma_k^2}{p_{\mathrm{T}}}\rho_k(t).
\end{equation}
 As a result, $\mathcal{P}_{2}$ can be reformulated as:
\begin{align}\label{sub-prob2-1}
&\underset{x_{m,n},\,u_k(t),
\rho_k(t),\mathbf F_o,\mathbf A_o,\theta_{\iota(n,m),o}}{\text{minimize}}
 s\,\Gamma^{\rm th}-\log\det\!\big(\mathbf I_{M}+s\,\mathbf R_\Sigma\,\mathbf Q\big)\nonumber\\
\text{s.t.}
&~\text{C1a},\nonumber~\text{C1b:} \ \sum_{t=1}^T u_k(t)\frac{\tau(t)}{T_{\max}}\log_2\big(1+\rho_k(t)\big)\ge R_{\min},\ \forall k,\\
&\text{C6:} \ \sum_{k=1}^K u_k(t)\le 1,\quad \text{C5:} \ u_k(t)\in\{0,1\},\ \forall k,t,\nonumber\\
&\text{C10}: \ \big(x_{m,n} - x_o\big)^2 
= \frac{1}{\text{Diag}(\mathbf{F}_o)_{\iota(n,m)}} - \hat{s}_{n,o}, \forall n,m,o,\nonumber\\ &\hat{s}_{n,o} \triangleq y_o^2 + d^2,\ \nonumber\\
&\text{C11}:\scalemath{0.9}{ \ [\mathbf{A}_o]_{\iota(n,m), \,\iota(n',m')} 
= e^{-j\big(\theta_{\iota(n,m),o}- \theta_{\iota(m',n'),o})\big)},\ \forall n,m,n',m',o,} \nonumber\\
&\text{C12}:\scalemath{0.9} {\theta_{\iota(n,m),o}
= \frac{2\pi}{\lambda} \left(\text{Diag}(\mathbf{F}_o)_{\iota(n,m)}\right)^{-\frac{1}{2}} 
+ \frac{2\pi}{\lambda_g}\,\ell_{m,n}}, \forall n,m,o, \nonumber\\ &\ell_{m,n} = \big\|\mathbf{x}_{m,n} - \mathbf{p}_{0}\big\|, \forall n,m, \nonumber\\
&\text{C13}: \ \mathbf{F}_o \succeq 0,\quad \mathbf{A}_o \succeq 0,\ \forall o, \nonumber\\
&\text{C14}: \ \text{rank}(\mathbf{F}_o) = 1,\quad \text{rank}(\mathbf{A}_o) = 1,\ \forall o, 
\end{align}
where $\iota(n,m)=(m-1)N_{\mathrm{T}}+n \in \{1, \ldots,N_\mathrm{tot}\}$, with $N_\mathrm{tot} = N_{\mathrm{T}}M$, maps indices from a two-dimensional to a single-dimensional stacked vector and $\text{Diag}(\mathbf{F}_{k})_{\iota{(n,m)}}=[\mathbf{F}_{k}]_{\iota(n,m),\iota_(n,m)}$ for simplicity of notation. We first address the non-convex equality constraints C10 and C12. Each can be written as a pair of equivalent inequalities. For C10, we have
\begin{align}
\text{C10a}:~&\scalemath{0.9}{(x_{m,n}-x_o)^2 - \tfrac{1}{\mathrm{Diag}(\mathbf F_o)_{\iota({n,m})}} + \hat s_{n,o} \le 0,\ \forall m,n,\ o\in\{k,e\}}, \\
\text{C10b}:~&\scalemath{0.9}{\tfrac{1}{\text{Diag}(\mathbf F_o)_{\iota({n,m})}} - \hat s_{n,o} - (x_{m,n}-x_o)^2 \le 0,\ \forall m,n,\ o\in\{k,e\}}.
\end{align}
 Likewise, C12 becomes
\begin{align}
\text{C12a}:~& \scalemath{0.9}{\tfrac{2\pi}{\lambda}\big(\text{Diag}(\mathbf F_o)_{\iota({n,m})}\big)^{-\frac{1}{2}} + \tfrac{2\pi}{\lambda_g}\,\ell_{m,n} - \theta_{\iota(n,m),o} \le 0}, \\
\text{C12b}:~& \scalemath{0.9}{\theta_{\iota(n,m),o} - \tfrac{2\pi}{\lambda}\big(\text{Diag}(\mathbf F_o)_{\iota({n,m})}\big)^{-\frac{1}{2}} - \tfrac{2\pi}{\lambda_g}\,\ell_{m,n} \le 0}.
\end{align}
Constraints C10 and C12 are replaced by two respective subconstraints: the first subconstraints (C10a and C12a) are convex, whereas the second subconstraints (C10b and C12b) are DC constraints.
\begin{lemma}
For any Hermitian $\mathbf X\succeq \mathbf 0$, $\operatorname{rank}(\mathbf X)=1$ if and only if $\|\mathbf X\|_*-\|\mathbf X\|_2=0$, where $\|\mathbf X\|_*$ denotes the trace norm \cite{Rank}. 

Utilizing Lemma 1, $\text{C14}$ can be equivalently written as
\end{lemma}
\begin{equation}
\overline{\text{C14}}:~\|\mathbf A_o\|_*-\|\mathbf A_o\|_2 \le 0,~\|\mathbf F_o\|_*-\|\mathbf F_o\|_2 \le 0,\ \ \forall o\in\{k,e\}.
\end{equation}

Without loss of generality, we parameterize $\mathbf A_k$ by its first nonzero column. Using trigonometric identities, $\text{C11}$ is equivalently expressed as
\begin{align}
\text{C11a}:~& \Re([\mathbf A_k]_{i,1})=\cos(\hat\theta_{i,k}),i \in \{2, \ldots, N_\mathrm{tot}\},\\
\text{C11b}:&\Im([\mathbf A_k]_{i,1})=-\sin(\hat\theta_{i,k}), i \in \{2, \ldots, N_\mathrm{tot}\},
\end{align}
where $\hat{\theta}_{i,k}=\theta_{i,k}-\theta_{1,k}$. To softly enforce constraints C11a and C11b, we add a quadratic penalty term to the objective function of the convex surrogate problem constructed using the MM framework: 
\begin{equation}
    \scalemath{0.9}{\Phi\!\left([\mathbf A_k]_{i,1},\hat\theta_{i,k}\right)
=\big|\Re([\mathbf A_k]_{i,1})-\cos\hat\theta_{i,k}\big|^2
+\big|\Im([\mathbf A_k]_{i,1})+\sin\hat\theta_{i,k}\big|^2},
\end{equation}
with a penalty coefficient $\varrho_{1}>0$, and minimize the penalized objective function subject to the remaining constraints. The resulting penalized problem reads
\begin{align}
&\underset{{x_{m,n},\,u_k(t),
\rho_k(t),\mathbf F_o,\mathbf A_o,\theta_{\iota(n,m),o}}}{\text{minimize}}
\hspace{-10.9mm} s\,\Gamma^{\rm th}-\log\det\!\big(\mathbf{I}_M+s\,\mathbf R_\Sigma\,\mathbf Q\big)
+\nonumber\\&\varrho_{1}\sum_{k=1}^{K}\sum_{i=2}^{N_{\rm tot}}
\Phi\!\left([\mathbf A_k]_{i,1},\hat\theta_{i,k}\right) \nonumber\\
\text{s.t.}\quad
&\text{C1},~\text{C5},~\text{C6},~\text{C10a}-\text{C12b},\ \text{C13},~\overline{\text{C14}}. \label{eq:penalizedP2}
\end{align}
Problem \eqref{eq:penalizedP2} and the original formulation in \eqref{sub-prob2-1} are equivalent when $\varrho_{1}$ is sufficiently large. Although \eqref{eq:penalizedP2} is still non-convex, it admits a standard MM approach \cite{MM}. In SCA iteration $l$, let $z_{m,n,k}=\text{Diag}(\mathbf F_k)_{\iota(n,m)}$ and define
\begin{equation}
\boldsymbol{\mathcal S}_{k}^{(l)}\triangleq\mathbf F_k^{(l)}+\mathbf C(\mathbf x^{(l)})^{\mathrm{H}}\mathbf A_k^{(l)}\mathbf C(\mathbf x^{(l)}).
\end{equation}
Using first-order Taylor surrogates in the current iteration $l$, we obtain the following global under-estimators for the DC terms:
\begin{align}
&\overline{\big\|\mathbf F_k+\mathbf C^{\mathrm{H}}\mathbf A_k\mathbf C\big\|_{\mathrm{F}}^2}
=\|\boldsymbol{\mathcal S}_{k}^{(l)}\|_{\mathrm{F}}^2
+2\,\text{Tr}\big((\boldsymbol{\mathcal S}_{k}^{(l)})^{\mathrm{H}}(\mathbf F_k-\mathbf F_k^{(l)})\big)\nonumber\\
&+2\,\text{Tr}\big((\boldsymbol{\mathcal S}_{k}^{(l)})^{\mathrm{H}}\mathbf C(\mathbf x^{(l)})^{\mathrm{H}}(\mathbf A_k-\mathbf A_k^{(l)})\mathbf C(\mathbf x^{(l)})\big),\\
&x^{\rm aff}_{m,n,k}=(x_{m,n}^{(l)}-x_k)^2+2(x_{m,n}^{(l)}-x_k)(x_{m,n}-x_{m,n}^{(l)}),\\
&z^{\rm aff}_{m,n,k}=\frac{1}{z^{(l)}_{m,n,k}}-\frac{z_{m,n,k}-z^{(l)}_{m,n,k}}{(z^{(l)}_{m,n,k})^2},\\&
\overline{z}^{\rm aff}_{m,n,k}=\frac{2\pi}{\lambda}(z^{(l)}_{m,n,k})^{-\frac12}-\frac{\pi}{\lambda}(z^{(l)}_{m,n,k})^{-\frac32}\big(z_{m,n,k}-z^{(l)}_{m,n,k}\big).
\end{align}
For the rank surrogates in $\overline{\text{C14}}$, we linearize the spectral norm at $\mathbf Y_o^{(l)}$ ($\mathbf Y\in\{\mathbf A,\mathbf F\}$):
\begin{align}
\|\mathbf Y_o\|_2  \ge \|\mathbf Y_o^{(l)}\|_2+\text{Tr}\!\Big(\bm\Phi_{\max}(\mathbf Y_o^{(l)})\bm\Phi^{\mathrm{H}}_{\max}(\mathbf Y_o^{(l)})\big(\mathbf Y_o-\mathbf Y_o^{(l)}\big)\Big),
\end{align}
where $\bm\Phi_{\max}(\cdot)$ denotes the principal right singular vector. Similarly, we apply the MM technique
along with a Lipschitz gradient surrogate to establish a 
global upper bound for $\Phi\!\left([\mathbf A_k]_{i,1},\hat\theta_{i,k}\right)$
\begin{align}
\overline{\Phi}^{(l)}_{i,k}
&= 2\!\left(\hat{\Re}_{i,k}^{(l)} \sin \hat{\theta}_{i,k}^{(l)} + \hat{\Im}_{i,k}^{(l)} \cos \hat{\theta}_{i,k}^{(l)}\right)\Delta \hat{\theta}_{i,k}+2\,\hat{\Re}_{i,k}^{(l)}\,\Delta \Re_{i,k}\ +\ \,\nonumber\\&2\hat{\Im}_{i,k}^{(l)}\,\Delta \Im_{i,k}
 +\ \frac{L_{\rm AR}}{2}\,(\Delta \Re_{i,k})^2\ +\ \frac{L_{\rm AI}}{2}\,(\Delta \Im_{i,k})^2\ +\nonumber\\&\frac{L_{\rm TH}}{2}\,(\Delta \hat{\theta}_{i,k})^2\ +\ \Phi^{(l)}_{i,k},
\end{align}
where $\hat{\Re}_{i,k}^{(l)}=\Re([\mathbf A_k]_{i,1}^{(l)})-\cos\hat{\theta}_{i,k}^{(l)}$, $\hat{\Im}_{i,k}^{(l)}=\Im([\mathbf A_k]_{i,1}^{(l)})+\sin\hat{\theta}_{i,k}^{(l)}$, $\Delta \hat{\theta}_{i,k}=\hat{\theta}_{i,k}-\hat{\theta}_{i,k}^{(l)}$, $\Delta \Re_{i,k}=\Re([\mathbf A_k]_{i,1})-\Re([\mathbf A_k]_{i,1}^{(l)})$, and $\Delta \Im_{i,k}=\Im([\mathbf A_k]_{i,1})-\Im([\mathbf A_k]_{i,1}^{(l)})$, with Lipschitz constants $L_{\rm AR}=L_{\rm AI}=2$ and $L_{\rm TH}=4$. In addition, since the mapping $\mathbf Q\mapsto\log\det(\mathbf I_M+s\mathbf R_\Sigma\mathbf Q)$ is concave in $\mathbf Q$, we linearize it at $\mathbf Q^{(l)}$:
\begin{align}
\log \det\left( \mathbf{I}_M + s\, \mathbf{R}_\Sigma \mathbf{Q} \right)
&\leq \log \det\left( \mathbf{A}^{(l)} \right) +\nonumber\\& 
\mathrm{Tr} \left( \left(\mathbf{A}^{(l)}\right)^{-1} s\, \mathbf{R}_\Sigma (\mathbf{Q} - \mathbf{Q}^{(l)}) \right),
\end{align}
which preserves the convexity of the MM subproblem. Collecting all surrogates, in the $(l{+}1)$-th SCA iteration, the following convex problem has to be solved. 
\begin{align}\label{BCD_Problem}
&\underset{x_{m,n},\,u_k(t),\,\rho_k(t),\,\mathbf F_o,\,\mathbf A_o,\,\theta_{\iota(n,m),o}}{\text{minimize}}
s\Gamma^{\rm th}
-\log \det\!\left(\mathbf A^{(l)}\right)
\nonumber\\
&
-\mathrm{Tr}\!\left(\left(\mathbf A^{(l)}\right)^{-1} s\,\mathbf R_\Sigma \bigl(\mathbf Q-\mathbf Q^{(l)}\bigr)\right)
+\varrho_{1}\sum_{k=1}^{K}\sum_{i=2}^{N_{\rm tot}}\overline{\Phi}^{(l)}_{i,k}
\nonumber\\
\text{s.t.}\quad
&\overline{\text{C1a}}:\;
\frac{\overline{\left\|\mathbf F_k + \mathbf C^{\mathrm H}\mathbf A_k \mathbf C \right\|_{\mathrm F}^2}}{2}
-\frac{\|\mathbf F_k\|_{\mathrm F}^2}{2}-
\nonumber\\
&\quad
\frac{\|\mathbf C^{\mathrm H}\mathbf A_k \mathbf C\|_{\mathrm F}^2}{2}
-\frac{N_T \sigma_k^2}{p_T}\,\rho_k(t)
\ge 0,\ \forall k,t,
\nonumber\\
&\overline{\text{C10a}}:
(x_{m,n}-x_o)^2 - z^{\rm aff}_{m,n,o} + \hat s_{n,o} \le 0, \forall  m,n,o,
\nonumber\\
&\overline{\text{C10b}}:
\frac{1}{\mathrm{Diag}(\mathbf F_o)_{\iota(n,m)}} - \hat s_{n,o} - x^{\rm aff}_{m,n,o} \le 0,
\ \forall m,n,o,
\nonumber\\
&\overline{\text{C12a}}:
\theta_{\iota(n,m),o}-\overline{z}^{\rm aff}_{m,n,o}-\frac{2\pi}{\lambda_g}\,\ell_{m,n}\le 0,\  \forall m,n,o,
\nonumber\\
&\overline{\text{C12b}}:\!\!
\scalemath{0.9}{
\frac{2\pi}{\lambda}\bigl(\mathrm{Diag}(\mathbf F_o)_{\iota(n,m)}\bigr)^{-1/2}\!\!\!+\frac{2\pi}{\lambda_g}\,\ell_{m,n}
-\theta_{\iota(n,m),o}
\le 0}, \forall m,n,o,
\nonumber\\
&\widehat{\text{C14}}:
\|\mathbf Y_o\|_*
-
\Bigl(
\|\mathbf Y_o^{(l)}\|_2
+
\mathrm{Tr}\!\bigl(
\bm\Phi_{\max}\bm\Phi_{\max}^{\mathrm H}
(\mathbf Y_o-\mathbf Y_o^{(l)})
\bigr)
\Bigr)
\le 0, 
\nonumber\\
&\qquad\qquad
\mathbf Y_o\in\{\mathbf A_o,\mathbf F_o\}, \forall o,
\nonumber\\
&\text{C1b},\text{C5},\text{C6},\text{C7}.
\end{align}
Here, $x^{\rm aff}_{m,n,o}$, $z^{\rm aff}_{m,n,o}$, and $\overline{z}^{\rm aff}_{m,n,o}$ denote the affine MM surrogates in the current iteration.

\subsection{Proposed BCD-Based Solution}

The reformulated problem in \eqref{BCD_Problem} remains challenging due to the coupling between optimization variables ${x}_{m,n}$,~$\mathbf{F}_o,~\mathbf{A}_o$, and the binary user scheduling variables $u_k(t)$. To efficiently solve it, we adopt a \emph{Block Coordinate Descent} (BCD) approach by partitioning the optimization variables into two subproblems that are alternately updated. The two subproblems are solved iteratively until convergence:
\begin{enumerate}
    \item \textbf{Step~1:} Solve Subproblem 2a via the MM approach to update $x_{m,n}$,~$\mathbf{F}_o, \mathbf{A}_o, \theta_{o,m,n}$, and $\rho_k(t)$.
    \item \textbf{Step~2:} Solve Subproblem 2b via a big-M-based convex problem with penalty factor to update $u_k(t)$ and $\rho_k(t)$.
\end{enumerate}

\subsubsection*{Subproblem 2a (PA Position and Amplitude--Phase Block)}

In this subproblem, we optimize $x_{m,n}, \mathbf{F}_o,\mathbf{A}_o$, $\theta_{o,m,n}$, and the auxiliary SNR variables $\rho_k(t)$ while fixing the user scheduling $u_k(t)$. The corresponding optimization problem is formulated as
\begin{align}\label{P2l}
&   \underset{{x_{m,n},\,
\rho_k(t),\mathbf F_o,\mathbf A_o,\theta_{\iota(n,m),o}}}{\text{minimize}}
\hspace{-1mm}s\Gamma^{\rm th}-\log\det(\mathbf I_M+s\mathbf R_\Sigma\mathbf Q^{(l)})
-\nonumber\\&\text{Tr}\!\Big(\mathbf G^{(l)}\big(\mathbf Q-\mathbf Q^{(l)}\big)\Big)
+\varrho_{1}\sum_{k=1}^{K}\sum_{i=2}^{N_{\rm tot}}\overline{\Phi}^{(l)}_{i,k} \nonumber\\
    \text{s.t.} \quad 
    & \text{C1a}, \notag \text{C1b}, \notag \text{C7}, \notag 
\overline{\text{C10a}},\overline{\text{C10b}},\overline{\text{C12a}},\overline{\text{C12b}}, \widehat{\text{C14}}, \notag \\
    & \mathbf{F}_o \succeq 0,~\mathbf{A}_o, \succeq 0, \quad \forall o.
\end{align}
This subproblem is convex and can be solved efficiently using CVX \cite{CVX,incremental_MM}.

\subsubsection*{Subproblem 2b (User Scheduling and QoS Block)}
In this subproblem, the PA positions and amplitude-phase variables are fixed to the values obtained in Subproblem 2a. We then optimize the user scheduling $u_k(t)$ and the communication SNR $\rho_k(t)$. To handle the binary nature of $u_k(t)$, we adopt a big-M formulation by introducing the auxiliary variable
    $\tilde{\rho}_k(t) \triangleq \rho_k(t) \, u_k(t)$,
while imposing the following constraints
\begin{align}
   & \text{C15}: 0 \le \tilde{\rho}_k(t) \le \rho_{k}(t), \\
   &\text{C16}: 0 \le \tilde{\rho}_k(t) \le \rho_{\max} u_k(t), \\
   & \text{C17}:\rho_{k}(t) - (1-u_k(t))\rho_{\max} \le \tilde{\rho}_k(t),
\end{align}
where $\rho_{\max} = 2^{R_{\min}} - 1$. As a result, the per-user QoS constraint in C1b becomes
\begin{equation}
  \widehat{\text{C1b}}:\sum_{t=1}^T \frac{\tau(t)}{T_{\max}} \log_2\!\big(1 + \tilde{\rho}_k(t)\big) \ge R_{\min}, 
    \quad \forall k.
\end{equation}
Next, we relax the scheduling variable as $0 \le u_k(t) \le 1$ and rewrite C5 as follows:
\begin{align}
&\text{C5a:}~0 \le u_k(t) \le 1,\\
&\text{C5b:}~\sum_{k,t} u_k(t)(1-u_k(t))\leq 0.
\end{align}
Constraint C5b is a DC function, and we use the first-order Taylor expansion to convert the non-convex constraint to the following convex constraint 
\begin{equation}
   \scalemath{0.9}{\widehat{\text{C5b:}}\!\sum_{k=1}^K \sum_{t=1}^T \Big(u_k^{(l)}(t)(1-u_k^{(l)}(t))\!\! +\!\!(1\!-\!2u_k^{(l)}(t))\big(u_k(t)-u_k^{(l)}(t)\big)\Big)\!\!\leq 0}.
\end{equation}
To promote binary scheduling, we introduce a penalty term with parameter $\varrho_{2}>0$. The resulting optimization problem in iteration $l$ is formulated as
\begin{align}\label{eq:lambda_2}
&\min_{\{u_k(t), \rho_k(t), \tilde{\rho}_k(t)\}}
s\Gamma^{\mathrm{th}}
-
\log \det\!\left(
\mathbf I_M
+
s \mathbf R_\Sigma \mathbf Q
\right)
+
\nonumber\\&\scalemath{0.9}{\hspace{-2mm}\varrho_{2}
\sum_{k=1}^K \sum_{t=1}^T \Big(u_k^{(l)}(t)(1-u_k^{(l)}(t))+(1-2u_k^{(l)}(t))\big(u_k(t)-u_k^{(l)}(t)\big)\Big)}\ \\
\text{s.t.} \quad 
    & \text{C1a}, \notag \widehat{\text{C1b}},\text{C5a},\text{C6},\text{C15}-\text{C17}.
\end{align}
This subproblem is convex and can be solved efficiently using CVX \cite{CVX,incremental_MM}.
\begin{algorithm}
\caption{MM-based Outage Minimization with Chernoff Bound}
\label{alg:bcd_isac}
\begin{algorithmic}[1]
\STATE \textbf{Input:} System parameters $\{M, N_\mathrm{T}, K, T_{\max}\}$, Chernoff grid $\mathcal{S} = \{s_1, \dots, s_S\}$, maximum number of iterations $L_{\max}$, and tolerance $\varepsilon$.
\STATE \textbf{Initialize:} Best objective value $\mathcal{F}^* \gets +\infty$.

\FORALL{$s \in \mathcal{S}$}
    \STATE \textbf{Initialize:} Cluster selection $b_m^{(0)}(t)$, time allocation $\tau^{(0)}(t)$, PA positions ${x}_{m,n}^{(0)}(t)$, user scheduling $u_k^{(0)}(t)$, SNR variables $\rho_k^{(0)}(t)$, and iteration index $l \gets 0$.
    
    \REPEAT
        \STATE \textbf{[Subproblem 1: Cluster Selection and Time Allocation]}\\
        \quad Fix $\{\mathbf{x}_{m,n}^{(l)}(t), u_k^{(l)}(t)\}$ and solve the convexified problem \eqref{P1} to update
        \[
            \{b_m^{(l+1)}(t), \tau^{(l+1)}(t)\}.
        \]
        \STATE \textbf{[Subproblem 2: Beamforming and User Scheduling]}\\
        \quad \textit{(a) PA Position and Beamforming Update:}\\
        \quad\quad For fixed $\{u_k^{(l)}(t)\}$, solve the MM–based convex subproblem \eqref{P2l} to update
        \[
            \{\mathbf{x}_{m,n}^{(l+1)}(t), \mathbf{F}_o^{(l+1)}, \mathbf{A}_o^{(l+1)}, \theta_{o,m,n}^{(l+1)}, \rho_k^{(l+1)}(t)\}.
        \]
        \quad \textit{(b) User Scheduling Update:}\\
        \quad\quad For fixed $\{\mathbf{x}_{m,n}^{(l+1)}(t), \mathbf{F}_o^{(l+1)}, \mathbf{A}_o^{(l+1)}, \theta_{o,m,n}^{(l+1)}\}$, 
        solve the big-M-based convex problem \eqref{eq:lambda_2} with MM-linearized binary penalty to update
        \[
            \{u_k^{(l+1)}(t), \rho_k^{(l+1)}(t)\}.
        \]
        \STATE \textbf{Update:} $l \gets l+1$.
    \UNTIL{Convergence, or $l=L_{\max}$
        \[
            \frac{|\mathcal{F}^{(l)}(s)-\mathcal{F}^{(l-1)}(s)|}{|\mathcal{F}^{(l-1)}(s)|}<\varepsilon,
        \]
        where $\mathcal{F}^{(l)}(s)$ is the Chernoff-bound objective in iteration $l$ for parameter $s$.
    }
    
    \STATE \textbf{Store best solution:}\\
    If $\mathcal{F}^{(l)}(s) < \mathcal{F}^*$, then set $\mathcal{F}^* \gets \mathcal{F}^{(l)}(s)$ and store $\mathcal{V}^* \gets \{b_m(t), \tau(t), \mathbf{x}_{m,n}(t), u_k(t), \rho_k(t)\}$.
\ENDFOR

\STATE \textbf{Output:} Optimized solution $\mathcal{V}^* = \{b_m^*(t), \tau^*(t), \mathbf{x}_{m,n}^*(t), u_k^*(t), \rho_k^*(t)\}$.
\end{algorithmic}
\end{algorithm}
\vspace{-5mm}
\subsection{Convergence Analysis}
For a fixed Chernoff parameter \( s \), the proposed algorithm updates three blocks of variables in an alternating fashion: \text{(i)} PA-cluster selection and time slot durations (\textbf{Subproblem 1} in \textbf{Algorithm \ref{alg:bcd_isac}}), \text{(ii)} intra-cluster PA antenna positions and their corresponding beamforming matrices (\textbf{Subproblem 2a} in \textbf{Algorithm \ref{alg:bcd_isac}}), and \text{(iii)} user scheduling along with the SNR auxiliary variables (\textbf{Subproblem 2b} in \textbf{Algorithm \ref{alg:bcd_isac}}). In each subproblem, one set of variables is updated while the others are held fixed. The convergence of the proposed alternating procedure follows from the MM principle applied to the Chernoff-bound surrogate problem. In each iteration, a convex surrogate that upper-bounds the main objective and coincides with it at the current point is minimized. This guarantees a non-increasing sequence of surrogate objective values. Since the surrogate objective is bounded from below, the sequence converges to a stationary point of the surrogate problem. Solving each convexified subproblem, therefore, yields a solution that does not increase the value of the surrogate objective. As a result, the sequence of objective values, denoted by \(\{\mathcal{F}^{(l)}(s)\}\) at iteration \(l\) for fixed \(s\), is monotonically non-increasing. Since the objective is bounded below (e.g., by zero in the case of the outage probability surrogate), the sequence \(\mathcal{F}^{(l)}(s)\) converges. Moreover, for sufficiently large penalty factors associated with the binary relaxation terms, i.e., $\varrho_{1}$ in \eqref{eq:penalizedP2}, $\varrho_{2}$ in \eqref{eq:lambda_2}, and $\rho$ in \eqref{P1}, the surrogate updates satisfy the regularity conditions required for convergence to a stationary point, as established in \cite{tseng2001convergence,khalili2025movable,khalili2024efficient}. Thus, each iteration improves or maintains the surrogate objective value, and the algorithm converges to a stationary solution of the surrogate problem for any fixed \(s\). The outer loop of the algorithm performs a one-dimensional search over a finite grid of values for the Chernoff parameter \(s\). For each candidate \(s\), the AO procedure is executed until convergence (see line 9 in \textbf{Algorithm \ref{alg:bcd_isac}}). Finally, the solution that yields the smallest surrogate objective value among all candidate \(s\) values is selected. This ensures a tractable and high-quality approximation to the original outage minimization problem \(\mathcal{P}_0\).
\vspace{-2mm}
\subsection{Complexity Analysis}
The overall complexity is dominated by the convex subproblems in each AO iteration and the outer search over the Chernoff parameter \( s \). Subproblem 1 scales as \( \mathcal{O}(MT) \), while Subproblem 2 is dominated by the semi-definite programming (SDP) in beamforming design with complexity \( \mathcal{O}((N_{\mathrm T}M)^3) \), and the user scheduling step with complexity \( \mathcal{O}(KT) \). Let \( S \) denote the number of grid points for \( s \) and \( I_{\mathrm{AO}} \) the number of AO iterations. Then, the total complexity is given by
\[
\mathcal{O}\!\bigg(S I_{\mathrm{AO}} \big((N_{\mathrm T}M)^3 + MT + KT \big)\bigg).
\]
\begin{table}[t]
\caption{Simulation Parameters}
\label{tab:simulation_parameters}
\centering
\setlength{\tabcolsep}{1pt}
\begin{tabular}{|c|c|c|c|}
\hline
\textbf{Parameter} & \textbf{Value} & \textbf{Parameter} & \textbf{Value} \\
\hline
Waveguide length ($D_x$) & 10 m & Number of pinching clusters ($M$) & 10 \\
Number of users & 2& Total time ($T_{\text{max}}$)& 8 ms\\
Receive antennas ($N_{\mathrm{R}}$) & 8 & SNR threshold ($\Gamma^{\text{th}}$) & 10 dB \\
Noise power ($\sigma_{k}^2=\sigma^2$) & –90 dBm & Average RCS power gain ($\zeta_{\mathrm{av}}$) & 1 m$^2$ \\
Carrier frequency ($f_c$) & 30 GHz & Attenuation factor ($\alpha$) & 0.18 \cite{docomo2019ntt} \\
Refractive index ($n_{\text{eff}}$) & 1.4 \cite{p3} & PA height ($d$) & 3 m \\
\hline
\end{tabular}
\end{table}
\vspace{-6mm}
\section{Simulation Results} 
In this section, we evaluate the performance of the proposed clustered PA-enabled ISAC design via computer simulation. We consider a square area of size $10 \times 10$~m$^2$ in the horizontal $xy$-plane. The DFRC-BS is positioned at $[0,\,0,\,d]$, where $d$ denotes the vertical distance between the waveguide and the area of interest. The PA clusters are deployed along the waveguide on the $x$-axis, with each cluster occupying a predefined section. Within each cluster, the PAs can move along the waveguide over a limited interval of length $L_c$, which defines the intra-cluster movement range. Specifically, the position of the $n$-th PA in cluster $m$ is optimized within this interval. The centers of the clusters are uniformly spaced along the waveguide, providing coverage of the entire area of interest. The default simulation parameters are summarized in Table~\ref{tab:simulation_parameters}. The proposed scheme is compared against four baseline designs. In baseline scheme~1, the reconfigurable PA architecture is replaced by a conventional transmit ULA consisting of $M N_{\mathrm{T}}$ antenna elements with uniform spacing of $\lambda/2$ along the $x$-axis and centered at $[5\, \text{m},\,0,\,d]$. The array is equally partitioned into $M$ subarrays with $N_\mathrm{T}$ elements to match the clustered structure of the proposed scheme. However, the antenna positions are fixed and are not optimized, and thus no intra-subarray antenna repositioning is performed. In baseline scheme~2, the same PA cluster is activated in all $T$ time slots. This baseline is obtained by solving problem $\mathcal{P}_{0}$ while enforcing the cluster selection variables to be identical across time, thereby disabling cluster switching while preserving intra-cluster antenna position optimization. In baseline scheme~3, all time slots are assigned equal durations. This baseline is obtained by solving problem $\mathcal{P}_{0}$, while fixing the time-allocation variables to $\tau(t)=\frac{T_{\max}}{T}$ for all $t$, thereby disabling adaptive time-slot duration optimization while retaining cluster selection, antenna positioning, and user scheduling. In baseline scheme~4, each PA cluster is equipped with a single antenna element. This baseline is obtained by solving problem 
$\mathcal{P}_{0}$, with the number of active antennas per cluster restricted to one, i.e., $N_\mathrm{T}=1$, thereby disabling intra-cluster beamforming while preserving cluster selection, time allocation, and user scheduling. For all schemes, we evaluate the performance based on the actual outage probability via Monte Carlo simulations.

\begin{figure}[t]\label{fig:outage_vs_threshold}
    \centering
\includegraphics[width=0.65\linewidth]{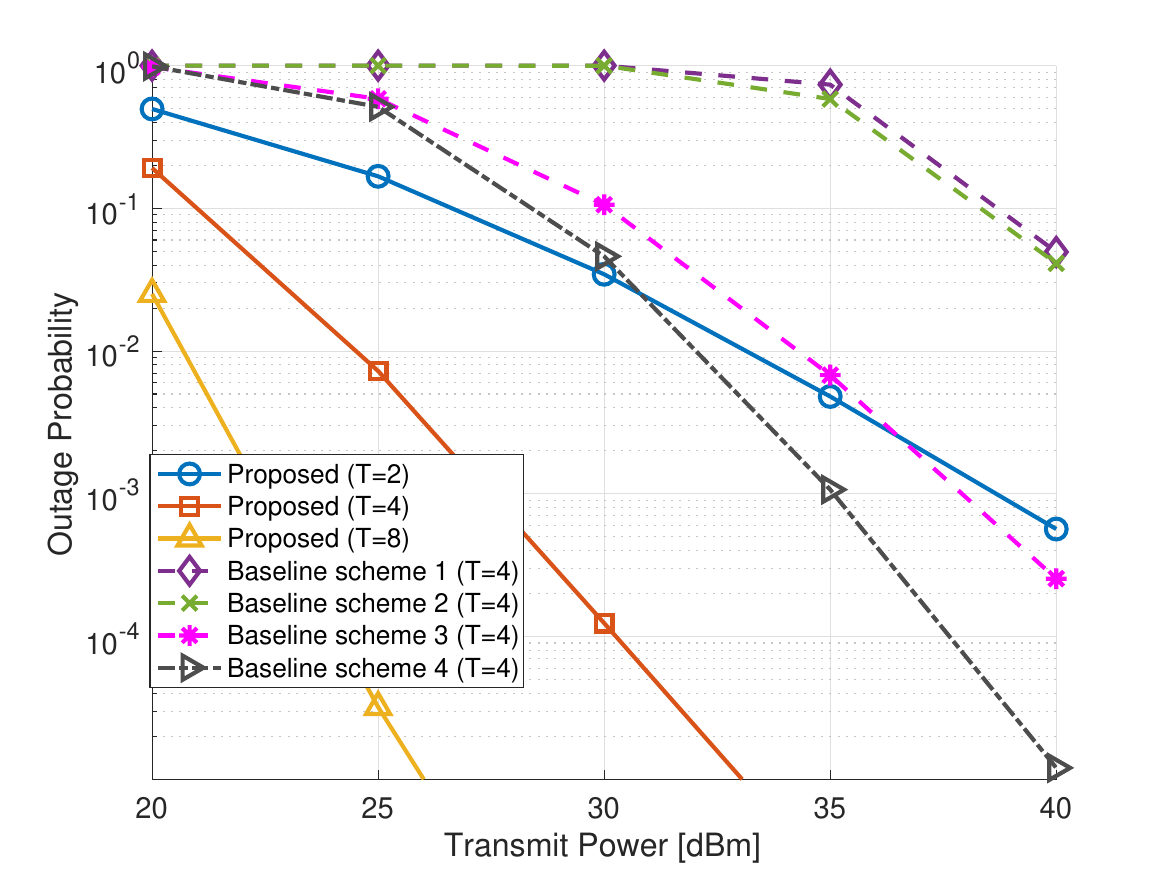}
\vspace{-2mm}
\caption{Outage probability versus transmit power for $R_{\min}=0.5$~bps/Hz, $N_\mathrm{T} = 4$, and $\kappa=0.1$.}
%\vspace{-5mm}
\end{figure}

Fig.~3 illustrates the sensing outage probability as a function of the transmit power for the proposed scheme and the four baseline schemes with  $R_\mathrm{min}=0.5$ bps/Hz and $\kappa =0.1$. In Fig.~3, baseline schemes~1-4 are evaluated for $T=4$, whereas the performance of the proposed scheme is shown for $T \in \{2,4,8\}$. As expected, increasing the transmit power improves the outage performance for all schemes, since stronger radar echoes enhance target detectability. For $T=4$, the proposed scheme achieves a lower outage probability than all baseline schemes. This performance gain results from the joint optimization of cluster selection, intra-cluster PA positioning, and cluster activation durations. By activating PA clusters at different locations along the waveguide, the proposed approach illuminates the target from multiple angular directions, thereby providing target diversity that improves robustness against angle-dependent RCS fluctuations. In contrast, neither baseline scheme~1 nor baseline scheme~2 exploit different spatial viewpoints over time, and hence both suffer from limited robustness against RCS variations induced by angle-dependent target scattering. Comparing baseline schemes~3 and~4, baseline scheme~4 consistently outperforms baseline scheme~3 for all considered transmit powers. Although both schemes exploit cluster switching across time slots, baseline scheme~3 operates with fixed slot durations and therefore cannot reallocate sensing time toward more favorable clusters. In contrast, baseline scheme~4, although it uses only a single PA per cluster and therefore does not provide an intra-cluster beamforming gain, it can benefit from flexible time allocation. A clear trend can also be observed with respect to the number of time slots $T$. As $T$ increases from $2$ to $8$, the system illuminates the target from additional spatial angles, effectively increasing the diversity order and improving sensing reliability. 

%Nevertheless, the proposed schemes with $T=4$ and $T=8$ consistently achieve the lowest outage probabilities overall. This confirms that combining multi-slot cluster activation with intra-cluster antenna-position optimization and adaptive time allocation provides the strongest sensing reliability, while simultaneously satisfying the communication QoS constraints of the scheduled users.

\begin{figure}[t]
    \centering
\includegraphics[width=0.65\linewidth]{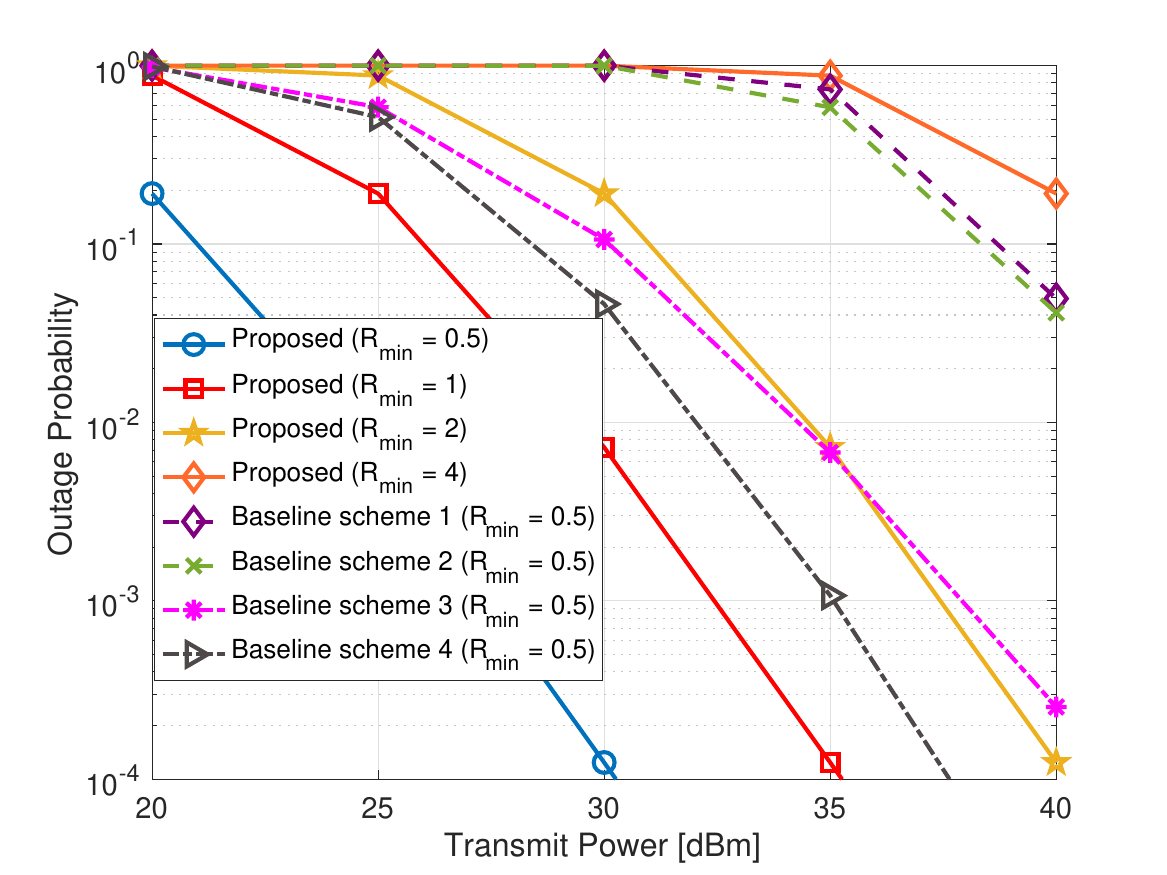}
\vspace{-4mm}
\caption{Outage probability versus transmit power for $T=4$, $N_\mathrm{T} = 4$, and $\kappa=0.1$.}
\end{figure}
 Fig.~4 depicts the sensing outage  probability as a function of the transmit power for $T=4$, $\kappa =0.1$, and different minimum communication rate requirements $R_{\min}$. For the proposed scheme, the outage probability increases as the minimum required QoS becomes more stringent. This behavior reflects the fundamental sensing-communication tradeoff: higher communication rate requirements restrict the feasible design space and reduce the possibilities for finding sensing-favorable cluster activations, antenna positions, and time-slot durations. This effect is particularly pronounced for $R_{\min}=4$~bps/Hz, where the proposed scheme exhibits the highest outage. In this case, the optimization prioritizes satisfying the communication constraints, which limits the flexibility to exploit angular diversity and allocate sensing time efficiently. As a result, the benefit of illuminating the target form different spatial angles is reduced, leading to degraded sensing reliability even at high transmit powers. For reference, Fig.~4 also shows the outage performance of the baseline schemes evaluated at $R_{\min}=0.5$~bps/Hz. We note that even for a moderately tighter requirement (e.g., $R_{\min}=1$~bps/Hz), the proposed scheme still achieves a lower outage probability than all baseline schemes for $R_{\min}=0.5$~bps/Hz, demonstrating that the joint optimization of cluster selection, intra-cluster antenna positioning, and slot durations can compensate the performance loss induced by stricter QoS constraints. 
 
 %Overall, Fig.~4 highlights the sensing-communication tradeoff and confirms that the proposed design remains highly effective as long as the communication constraint is not overly restrictive.

\begin{figure}[t]
    \centering
\includegraphics[width=0.65\linewidth]{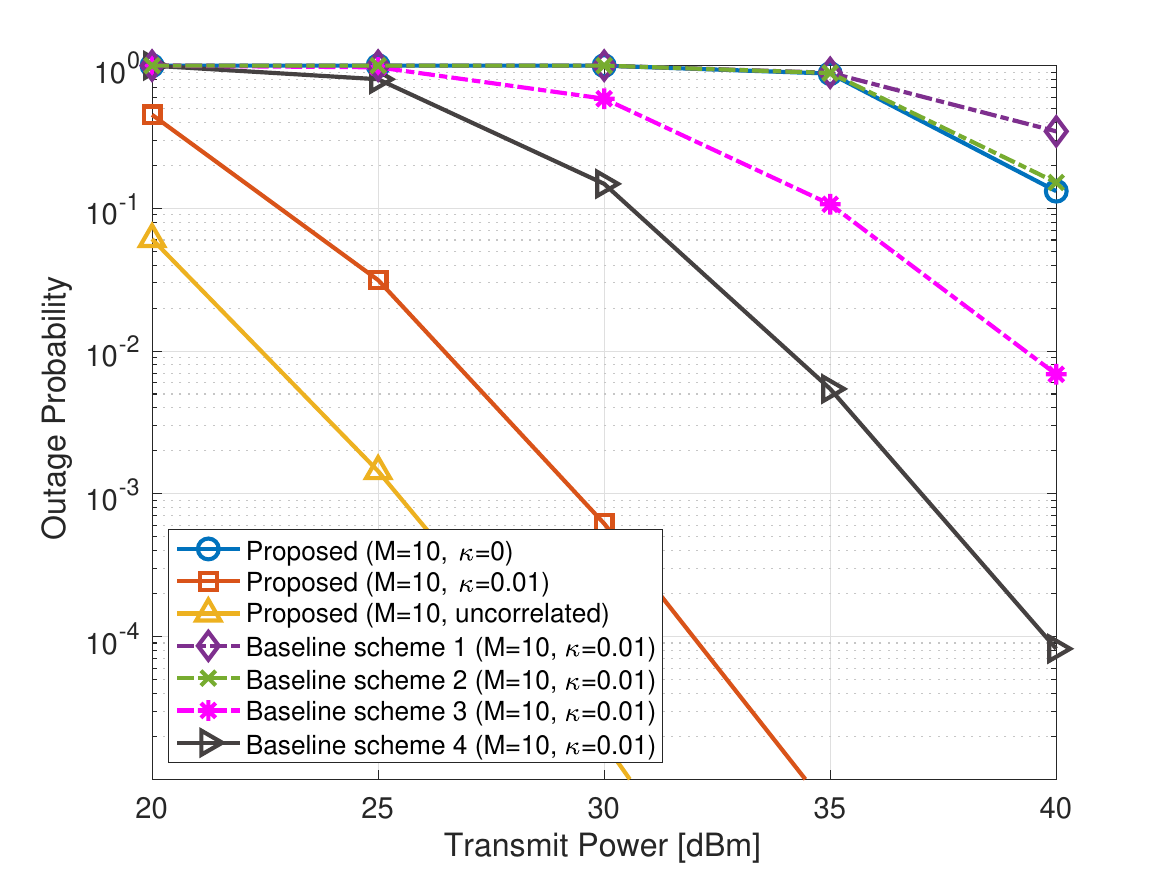}
\vspace{-4mm}
\caption{Outage probability versus transmit power for $T=4$, $N_\mathrm{T} = 4$, and $R_{\min}=0.5$ bps/Hz.}
\end{figure}
Fig.~5 illustrates the sensing outage  probability as a function of the transmit power for different RCS correlations. Three correlation levels are considered for the proposed scheme for $T=4$, $N_\mathrm{T} = 4$, and $R_{\min}=0.5$: fully correlated RCS ($\kappa = 0$), partially correlated RCS ($\kappa = 0.01$), and uncorrelated RCS ($\kappa\rightarrow \infty$). The fully correlated case represents the most adverse scenario, where all PA positions experience identical RCS fluctuations and spatial diversity across different look angles is effectively lost. As a result, the performance of the proposed scheme with $\kappa = 0$ is similar to that of baseline scheme 2, since angular diversity is completely lost and antenna reconfigurability no longer provides a sensing advantage. This observation suggests that the benefits of PA reconfigurability critically rely on angular variability in the RCS. As the RCS correlation decreases, i.e., as $\kappa$ increases, the sensing performance of the proposed scheme improves significantly. In the partially correlated case ($\kappa = 0.01$), activated different PA clusters experience partially decorrelated RCS realizations across time slots, enabling the system to exploit target diversity and achieve noticeably lower outage probabilities. This performance gain becomes even more pronounced in the uncorrelated scenario, where the proposed scheme fully leverages spatial and temporal reconfigurability to accumulate sensing energy from multiple statistically independent look angles.

%Overall, Fig.~5 demonstrates that the performance gains offered by PAs are fundamentally tied to the presence of angular-dependent RCS fluctuations. When such variability exists, the proposed framework can effectively harness both spatial and temporal diversity together with intra cluster beamforming to substantially reduce sensing outage  probability. Conversely, when the RCS is fully correlated across angles, the advantage of PA cluster reconfigurability diminishes, and the system performance converges toward that of conventional static antenna architectures.

Fig.~6 illustrates the sensing outage  probability as a function of the transmit power for different numbers of PAs per cluster $N_{\mathrm{T}}$, with $T=4$, $\kappa =0.1$, and $R_{\min}=0.5$~bps/Hz. For the proposed scheme, increasing $N_{\mathrm{T}}$ from $2$ to $6$ leads to a consistent and substantial reduction in outage probability across the entire transmit power range. This trend confirms that adding more PAs within each cluster enhances the beamforming capability, thereby increasing the effective sensing gain and improving robustness against RCS fluctuations. To further highlight the role of beamforming, Fig.~6 also includes the performance of the proposed scheme without enforcing intra-cluster beamforming for $N_{\mathrm{T}}=4$. Specifically, we solve $\mathcal{P}_0$ with fixed PA positions within the active cluster, i.e., without solving Subproblem 2a. As a result, the transmitted signals may add partially destructively, leading to a reduced effective sensing gain compared to the beamforming-based implementation. This performance gap demonstrates that the sensing gains of the PAs do not arise merely from increasing the number of radiating elements, but critically rely on proper phase alignment achieved through intra-cluster beamforming. An additional insight from Fig.~6 is that baseline scheme~4, which employs a single antenna per cluster, can outperform the proposed scheme without beamforming, despite using fewer radiating elements. This is because, without phase alignment, the signal transmitted by multiple antenna within a cluster are not guaranteed to add constructively, which negatively affects the outage performance. This observation confirms that simply increasing the number of active antennas is insufficient; appropriate phase alignment is essential to fully exploit the sensing potential of clustered PAs.

\begin{figure}[t]
    \centering
\includegraphics[width=0.65\linewidth]{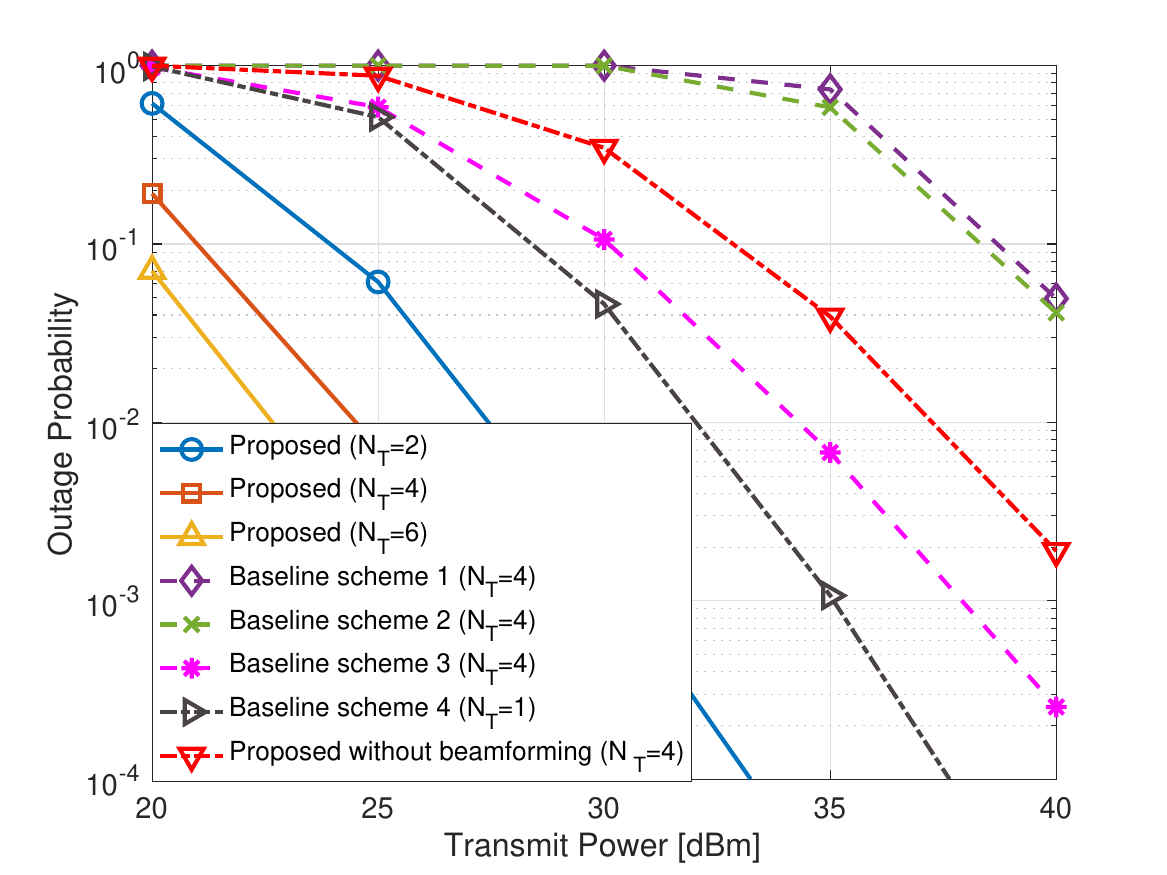}
\vspace{-4mm}
\caption{Outage probability versus transmit power for $T=4$, $R_{\min}=0.5$ bps/Hz, and $\kappa=0.1$.}
\end{figure}

\begin{figure*}[t]
\centering

\begin{minipage}[t]{0.5\textwidth}
  \centering
  \includegraphics[width=\linewidth]{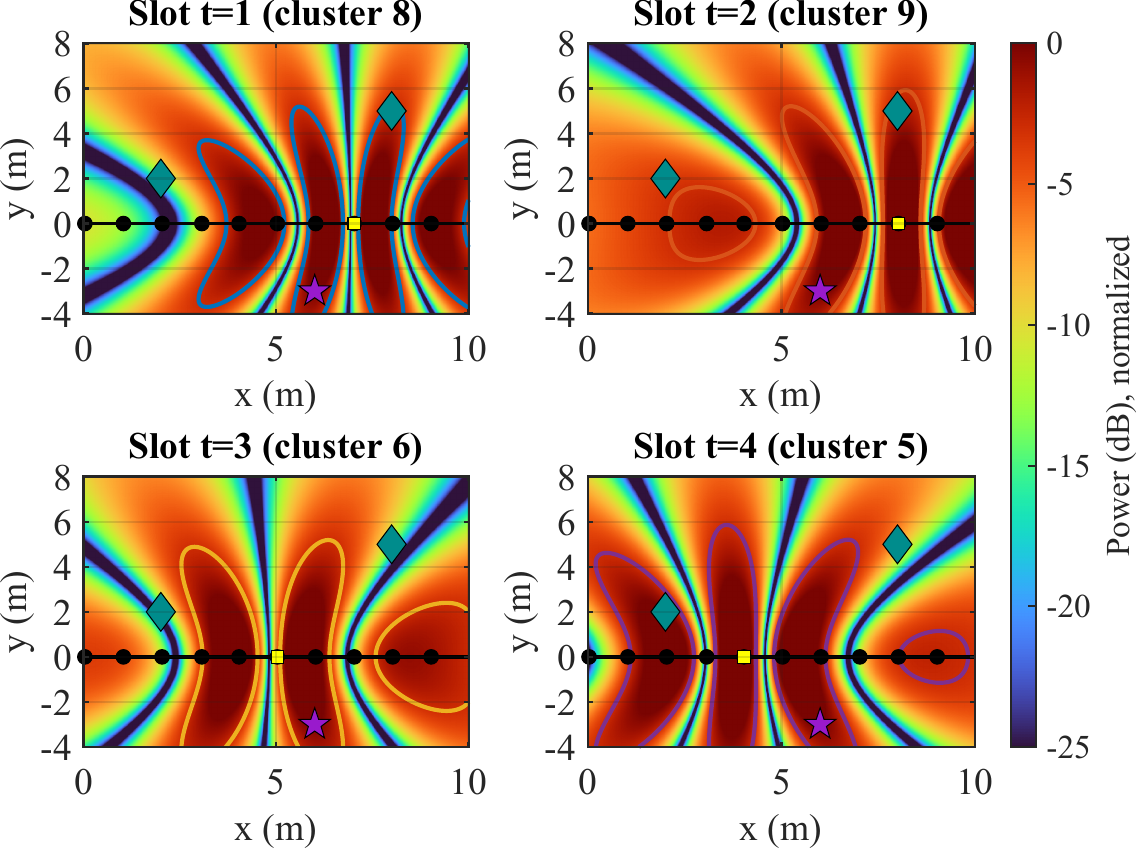}\\[-1.5mm]
  {\footnotesize (a) Proposed scheme}
\end{minipage}\hfill
\begin{minipage}[t]{0.5\textwidth}
  \centering
  \includegraphics[width=\linewidth]{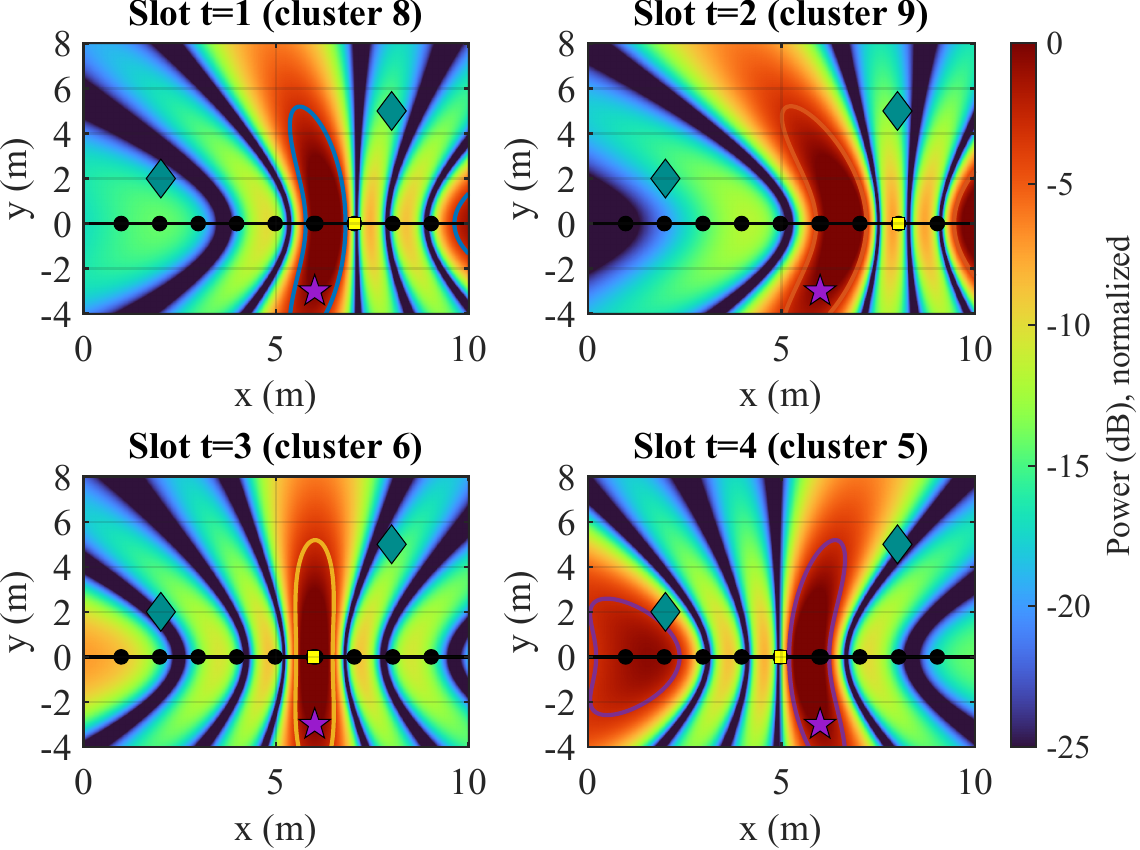}\\[-1.5mm]
  {\footnotesize (b) Target-aligned baseline}
\end{minipage}\hfill
\begin{minipage}[t]{0.5\textwidth}
  \centering
  \includegraphics[width=\linewidth]{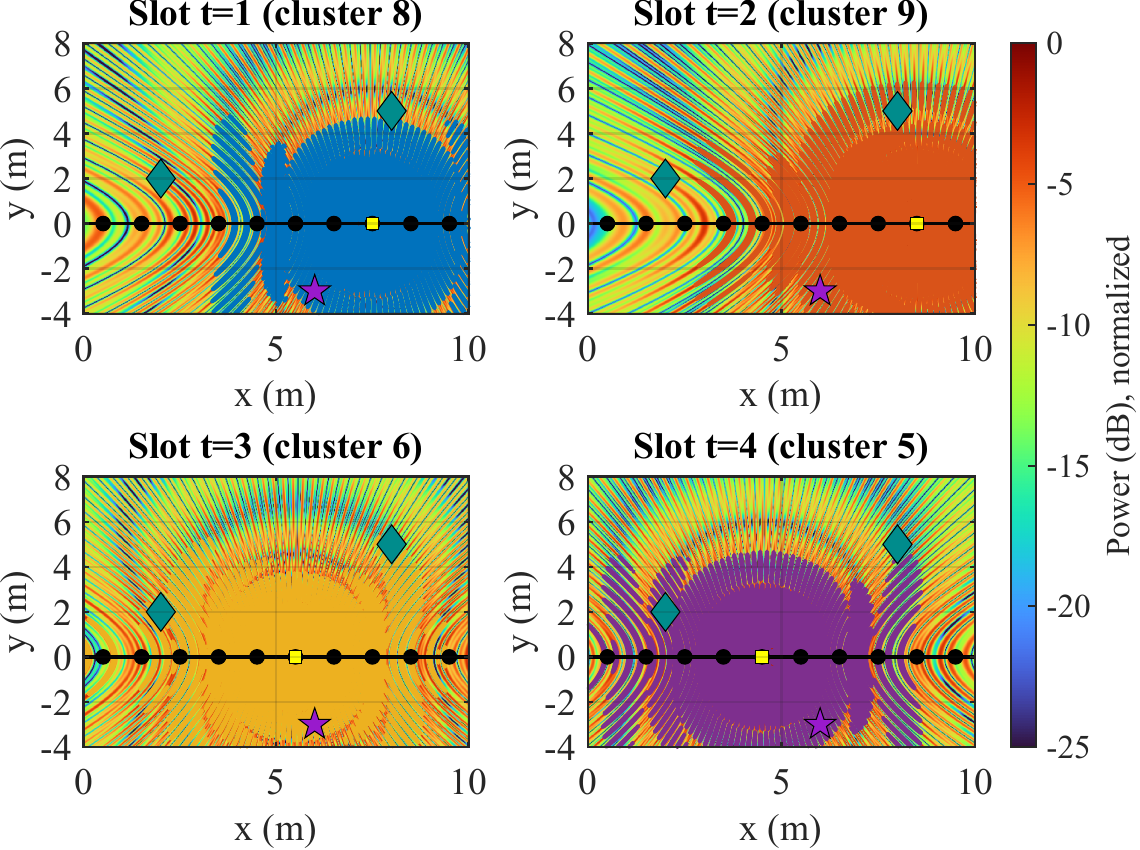}\\[-0.1mm]
  {\footnotesize (c) Uniform baseline}
\end{minipage}

\vspace{0.5mm}
\includegraphics[width=0.6\textwidth]{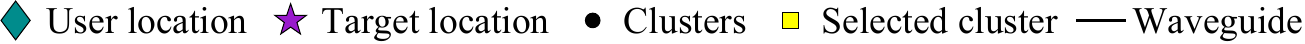}

\vspace{-2mm}
\caption{Spatial distribution of normalized radiated sensing power for different antenna positioning strategies with $T=4$, $R_{\min}=0.5$ bps/Hz, and $\kappa=0.1$. (a) Proposed scheme with optimized antenna positions in each active cluster. (b) Target-aligned baseline, where antenna positions are chosen based on geometric alignment with the target. (c) Uniform baseline, where antennas are evenly spaced within each cluster.}\label{fig:fig7}
\end{figure*}

Fig.~7 illustrates the spatial distribution of the normalized radiated sensing power (in dB) over $T=4$ time slots for three antenna positioning strategies. In each time slot, only one PA cluster is active, and the plotted contours show the resulting transmit power distribution generated by the antennas within the selected cluster. For the convenience of illustration, the power is normalized to its maximum value. The waveguide, PA clusters, target, and communication users are shown for reference. Fig.~7(a) corresponds to the proposed antenna positioning scheme. In each time slot, the antenna positions within the active cluster are jointly optimized to satisfy both sensing and communication requirements. Consequently, the transmitted power is effectively directed toward the target while maintaining sufficient coverage toward the communication users. Furthermore, since different clusters are activated across time slots, the target is illuminated from multiple spatial perspectives, which improves sensing robustness. Nevertheless, the achievable focusing remains constrained by the limited repositioning range of the antennas within each cluster. The structured ripple-like patterns observed in Fig.~7(a) arise from the distance-dependent phase differences of the spherical waves radiated by the antennas within the active cluster. Specifically, the phase of the signal radiated by each antenna varies proportionally to its propagation distance to each spatial location. As a result, the superposition of the radiated signals produces alternating regions of constructive and destructive interference. The spacing and orientation of these patterns depend on the antenna positions, the wavelength, and the relative geometry between the antennas and observation points. Since the antenna positions are optimized in the proposed scheme, the resulting interference patterns are shaped such that constructive superposition occurs near the target and communication users, while destructive interference appears elsewhere. This leads to the quasi-periodic contours visible in Fig.~7(a), which reflect the physical wave propagation and the spatial phase structure induced by the optimized antenna configuration. Fig.~7(b) shows the target-aligned baseline, where antenna positions are selected solely based on geometric alignment with the target. While this strategy concentrates the radiated power toward the target region, it does not account for the communication users, resulting in weaker coverage at their locations compared to the proposed scheme. Fig.~7(c) presents the uniform placement baseline, where antennas are evenly spaced within each cluster, and their positions remain fixed. In this case, the transmitted power is broadly distributed across the environment, resulting in weaker target illumination and less efficient use of the available transmit power. 
\vspace{-5mm}
\section{Conclusion}
\vspace{-1mm}
This paper presented a novel PA-enabled ISAC framework, where dynamically activatable PA clusters are distributed along a dielectric waveguide to facilitate multi-perspective target illumination. To capture realistic sensing behavior, we introduced a correlated complex Gaussian RCS model, where RCS realizations associated with nearby illumination angles are statistically correlated, reflecting the smooth variation of the target’s scattering response across look angles. Based on this model, we defined the sensing outage probability as a reliability metric and formulated an optimization problem for its minimization by jointly selecting the active PA clusters, the intra-cluster PA positions for transmit beamforming, and the cluster activation durations, subject to average data rate constraints for the communication users. To tackle the inherent non-convexity of the problem, we derived a tractable Chernoff-bound-based surrogate objective function and proposed an AO algorithm leveraging MM and penalty methods. Simulation results demonstrated that the proposed framework consistently outperforms the considered baseline schemes and provided several key insights. First, sequential activation of distinct PA clusters across time slots enables target diversity and leads to a substantial reduction in sensing outage probability compared to baseline schemes that either employ fixed antenna configurations, repeatedly activate the same PA cluster across all time slots, or rely on single PA clusters. Second, the benefits of PA cluster reconfigurability persist under moderately correlated RCS realizations, whereas fully correlated RCSs limit target diversity. Third, more stringent communication QoS requirements constrain sensing performance, revealing a fundamental trade-off between communication reliability and sensing robustness in PA-assisted ISAC systems. Finally, intra-cluster beamforming is essential to fully exploit the spatial degrees of freedom enabled by PA position optimization, whereas merely increasing the number of PAs per cluster without optimizing their positions to achieve constructive signal superposition at the target does not necessarily improve sensing performance.
\vspace{-5mm}
\bibliographystyle{IEEEtran}
\bibliography{Ref.bib}

@article{zhu2023movable,
  author={Zhu, Lipeng and Ma, Wenyan and Ning, Boyu and Zhang, Rui},
  journal={IEEE Trans. Wireless Commun.}, 
  title={Movable-Antenna Enhanced Multiuser Communication via Antenna Position Optimization}, 
year={Jul. 2024},
  volume={23},
  number={7},
  pages={7214-7229},
  doi={10.1109/TWC.2023.3338626}}

@article{Radar,
  title={Introduction to radar},
  author={Skolnik, Merrill I},
  journal={Radar {H}andbook},
  volume={2},
  pages={21},
  year={1962},
  publisher={McGraw-Hill New York, NY, USA}
}

@article{ISAC6G,
  title={Integrated Sensing and Communications: Toward Dual-Functional Wireless Networks for {6G} and Beyond}, 
	author={Liu, Fan and others},
  journal={IEEE J. Sel. Areas Commun.}, 
  year={Jun. 2022},
  volume={40},
  number={6},
  pages={1728-1767},
  doi={10.1109/JSAC.2022.3156632}}

@ARTICLE{mu-mimo-jsc,
	author={Liu, Fan and others},
	journal={IEEE Trans. Wireless Commun.}, 
	title={{MU-MIMO} Communications With {MIMO} Radar: From Co-Existence to Joint Transmission}, 
	year={Apr. 2018},
	volume={17},
	number={4},
	pages={2755-2770},
	doi={10.1109/TWC.2018.2803045}}

@ARTICLE{jsc-mimo-radar,
	author={Liu, Xiang and others},
	journal={IEEE Trans. Signal Process.}, 
	title={Joint Transmit Beamforming for Multiuser {MIMO} Communications and {MIMO} Radar}, 
	year={Jun. 2020},
	volume={68},
	number={},
	pages={3929-3944},
	doi={10.1109/TSP.2020.3004739}}

@article{tseng2001convergence,
  title={Convergence of a block coordinate descent method for nondifferentiable minimization},
  author={Tseng, Paul},
  journal={Journal of Optimization Theory and Applications},
  volume={109},
  number={3},
  pages={475--494},
  year={2001},
  month={Jun.},
  publisher={Springer}
}

@article{khalili2024efficient,
author={Khalili, Ata and others},
  journal={IEEE Trans. Wireless Commun.}, 
  title={Efficient {UAV} Hovering, Resource Allocation, and Trajectory Design for {ISAC} With Limited Backhaul Capacity},
  year={Nov. 2024},
  volume={23},
  number={11},
  pages={17635-17650},
  }

@article{wong2020fluid,
  title={Fluid antenna systems},
  author={Wong, Kai-Kit and Shojaeifard, Arman and Tong, Kin-Fai and Zhang, Yangyang},
  journal={IEEE Trans. Wireless Commun.},
  volume={20},
  number={3},
  pages={1950--1962},
  year={Mar. 2021},
  publisher={IEEE}
}

@article{docomo2019ntt,
  title={Pinching antenna: Using a dielectric waveguide as an antenna},
  author={Suzuki, Hiroshi Okazaki Yasunori and Kawai, Kunihiro},
  journal={NTT DOCOMO Technical J.},
  year={2022}
}

@ARTICLE{p1,
  author={Ding, Zhiguo and Schober, Robert and Vincent Poor, H.},
  journal={IEEE Trans Commun.}, 
  title={Flexible-Antenna Systems: A Pinching-Antenna Perspective}, 
  year={Oct. 2025},
  volume={73},
  number={10},
  pages={9236-9253},
  doi={10.1109/TCOMM.2025.3555866}}

@article{p3,
  author={Wang, Kaidi and Ding, Zhiguo and Schober, Robert},
  journal={IEEE Wireless Commun. Lett.}, 
  title={Antenna Activation for {NOMA} Assisted Pinching-Antenna Systems}, 
  year={May 2025},
  volume={14},
  number={5},
  pages={1526-1530},
  doi={10.1109/LWC.2025.3548280}}

@article{p4,
    author={Tegos, Sotiris A. and Diamantoulakis, Panagiotis D. and Ding, Zhiguo and Karagiannidis, George K.},
  journal={IEEE Wireless Commun. Lett.}, 
  title={Minimum Data Rate Maximization for Uplink Pinching-Antenna Systems}, 
  year={May 2025},
  volume={14},
  number={5},
  pages={1516-1520},
  doi={10.1109/LWC.2025.3547956}}

@ARTICLE{Ang1,
  author={Haimovich, Alexander and others},
  journal={IEEE Signal Process. Mag.}, 
  title={{MIMO} Radar with Widely Separated Antennas}, 
  year={Dec. 2008},
  volume={25},
  number={1},
  pages={116-129},
  doi={10.1109/MSP.2008.4408448}}

@INPROCEEDINGS{Ang2,
  author={Fishler, E. and others},
  booktitle={Proc. IEEE Asilomar Conference on Signals, Systems and Computers}, 
  title={Performance of {MIMO} radar systems: advantages of angular diversity}, 
  year={2004},
  pages={305-309},
  doi={10.1109/ACSSC.2004.1399142}}

@INPROCEEDINGS{khalili2025pinching,
  author={Khalili, Ata and Kaziu, Brikena and Papanikolaou, Vasilis K. and Schober, Robert},
  booktitle={Proc. IEEE GLOBECOM}, 
  title={Pinching Antenna-enabled ISAC Systems: Exploiting Look-Angle Dependence of RCS for Target Diversity}, 
  year={2025},
  volume={},
  number={},
  pages={387-392},
  doi={10.1109/GLOBECOM59602.2025.11432471}}

@ARTICLE{Exp,
  author={Cho, Hoonkyung and Chun, Joohwan and Lee, Taeseung and Lee, Seungjae and Chae, Daeyoung},
  journal={IEEE Trans. Aerosp. Electron. Syst.}, 
  title={Spatiotemporal radar target identification using radar cross-section modeling and hidden Markov models}, 
  year={Jul. 2016},
  volume={52},
  number={3},
  pages={1284-1295},
  doi={10.1109/TAES.2016.140908}}

@book{richards2014fundamentals,
  title={Fundamentals of Radar Signal Processing},
  author={Richards, Mark A.},
  publisher={McGraw-Hill Education},
  year={2014}
}

@article{PASS,
  title     = {Integrated Sensing and Communications for Pinching-Antenna Systems ({PASS})},
  author    = {Zheng Zhang and Yuanwei Liu and Bingtao He and Jian Chen},
  journal   = {arXiv preprint arXiv:2504.07709},
  year      = {2025},
  url       = {https://arxiv.org/abs/2504.07709}
}

@article{PA_ISAC_sensing,
  title     = {Pinching Antenna System for Integrated Sensing and Communications},
  author    = {Haochen Li and Ruikang Zhong and Jiayi Lei and Yuanwei Liu},
  journal   = {arXiv preprint arXiv:2508.19540},
  year      = {2025},
  url       = {https://arxiv.org/abs/2508.19540}
}

@article{PA_ISAC_rate,
  title     = {Rate Region of {ISAC} for Pinching-Antenna Systems},
  author    = {Chongjun Ouyang and Zhaolin Wang and Yuanwei Liu and Zhiguo Ding},
  journal   = {arXiv preprint arXiv:2505.10179},
  year      = {2025},
  url       = {https://arxiv.org/abs/2505.10179}
}

@INPROCEEDINGS{PA_ISAC_CRB,
  author    = {Dimitrios Bozanis and Vasilis K. Papanikolaou and Sotiris A. Tegos and George K. Karagiannidis},
  booktitle={Proc. IEEE PIMRC}, 
  title     = {Cram{\'e}r--{R}ao Bounds for Integrated Sensing and Communications in Pinching-Antenna Systems},
year={2025},
  volume={},
  number={},
  pages={1-6},
}

@article{PA_multi_waveguide,
  title     = {Multi-Waveguide Pinching Antennas for {ISAC}},
  author    = {W. Mao and others},
  journal   = {arXiv preprint arXiv:2505.24307},
  year      = {2025},
  url       = {https://arxiv.org/abs/2505.24307}
}

@ARTICLE{khalili2025movable,
  author={Khalili, Ata and Schober, Robert},
  journal={IEEE Trans. Wireless Commun.}, 
  title={Movable Antenna-Enabled {ISAC}: Tackling Slow Antenna Movement, Dynamic {RCS}, and Imperfect {CSI} via Two-Timescale Optimization}, 
  year={Jan. 2026},
  volume={25},
  number={},
  pages={8452-8467},
  doi={10.1109/TWC.2025.3638255}}

@ARTICLE{Correlation,
  author={Zhang, Wence and Ren, Hong and Pan, Cunhua and Chen, Ming and de Lamare, Rodrigo C. and Du, Bo and Dai, Jianxin},
  journal={IEEE Trans. Commun.}, 
  title={Large-Scale Antenna Systems With {UL}/{DL} Hardware Mismatch: Achievable Rates Analysis and Calibration}, 
  year={Apr. 2015},
  volume={63},
  number={4},
  pages={1216-1229},
  doi={10.1109/TCOMM.2015.2395432}}

@ARTICLE{MM,
  author={Sun, Ying and Babu, Prabhu and Palomar, Daniel P.},
  journal={IEEE Trans Signal Process.}, 
  title={Majorization-Minimization Algorithms in Signal Processing, Communications, and Machine Learning}, 
  year={Feb. 2017},
  volume={65},
  number={3},
  pages={794-816},
  doi={10.1109/TSP.2016.2601299}}

@article{incremental_MM,
  title={Incremental majorization-minimization optimization with application to large-scale machine learning},
  author={Mairal, Julien},
  journal={SIAM Journal on Optimization},
  volume={25},
  number={2},
  pages={829--855},
  year={2015},
  publisher={SIAM}
}

@misc{cvx,
  author       = {Michael Grant and Stephen Boyd},
  title        = {{CVX}: Matlab Software for Disciplined Convex Programming, Version 2.1},
  year         = {2014},
  month        = mar,
  howpublished = {[Online]. Available: http://cvxr.com/cvx}
}

@article{chernoff,
  title={A measure of asymptotic efficiency for tests of a hypothesis based on the sum of observations},
  author={Chernoff, Herman},
  journal={The Annals of Mathematical Statistics},
  pages={493--507},
  year={1952},
  publisher={JSTOR}
}

@inproceedings{Rank,
  author    = {M. Fazel and H. Hindi and S. Boyd},
  title     = {A rank minimization heuristic with application to minimum order system approximation},
  booktitle = {Proceedings of the American Control Conference (ACC)},
  volume    = {6},
  pages     = {4734--4739},
    year      = {2001}
}
\end{document}